\documentclass[twocolumn, prc, amssymb, superscriptaddress,aps,preprintnumbers,amsmath,floatfix]{revtex4}
\usepackage{multirow}
\usepackage{graphicx}% Include figure files
\usepackage{dcolumn}% Align table columns on decimal point
\usepackage{bm}% bold math
\usepackage{color}
\usepackage{amsmath}%[fleqn]
\usepackage{wasysym}
\usepackage{float}
\usepackage{array,url,kantlipsum}
\usepackage{verbatim}
\usepackage{xcolor}
\usepackage{siunitx}
\usepackage{CJK}
\usepackage{dsfont}
\usepackage{textcomp}
\newcolumntype{P}[1]{>{\centering\arraybackslash}p{#1}}
\usepackage{array}
\usepackage{braket}
\usepackage{lipsum}
\usepackage{upgreek}

\newcommand{\lrb}[1]{\left(#1 \right)} % saves time, lrb stands for left right bracket () 
\newcommand{\lrsb}[1]{\left [#1 \right]} %lrsb stands for left right squared bracket [] 
\newcommand{\lrcb}[1]{\left \{#1 \right \}}
\newcommand{\ab}[1]{\left |#1 \right |}
\begin{document}

%\title{Nuclear photoabsorption in $^{229}$Th using twisted light}

\title{Photoexcitation of the  $^{229}$Th nuclear clock transition using twisted light}

%%%%%%%%%%%%%%%%%%%%%%%%%%%%%%%%%%%%%%%%%%%%%%%%%%%%%%%%%%%%%%%%%%%%%%%%%%%%%%%%%

\author{Tobias \surname{Kirschbaum}}
\email{tobias.kirschbaum@uni-wuerzburg.de}
\affiliation{University of W\"urzburg, Institute of Theoretical Physics and Astrophysics,  Am Hubland, 97074 W\"urzburg, Germany}

\author{Thorsten Schumm}
\affiliation{Institute for Atomic and Subatomic Physics, TU Wien, Stadionallee 2, 1020 Vienna, Austria}

\author{Adriana P\'alffy}
\email{adriana.palffy-buss@uni-wuerzburg.de}
%\affiliation{ Max-Planck-Institut f\"ur Kernphysik, Saupfercheckweg 1, 69117 Heidelberg, Germany}
\affiliation{University of W\"urzburg, Institute of Theoretical Physics and Astrophysics,  Am Hubland, 97074 W\"urzburg, Germany}

%%%%%%%%%%%%%%%%%%%%%%%%%%%%%%%%%%%%%%%%%%%%%%%%%%%%%%%%%%%%%%%%%%%%%%%%%%%%%%%%%

\date{\today}

\begin{abstract}

The $^{229}$Th nucleus has a unique transition at only 8 eV which could be used for a novel nuclear clock. We investigate theoretically the prospects of driving this transition with vortex light beams carrying orbital angular momentum. Numerical results are presented for two experimental configurations which are promising for the design of the planned nuclear clock: a trapped ion setup and a large ensemble of nuclei doped into CaF$_2$ crystals
which are transparent in the frequency range of the nuclear transition. We discuss the feasibility of the vortex beam nuclear excitation and compare the excitation features with the case of plane wave beams. 

\end{abstract}
%%%%%%%%%%%%%%%%%%%%%%%%%%%%%%%%%%%%%%%%%%%%%%%%%%%%%%%%%%%%%%%%%%%%%%%%%%%%%%%%%

\maketitle

\section{Introduction}
\label{intro}

Twisted light or optical vortex beams refer to light beams with engineered wave fronts that carry orbital angular momentum along their direction of propagation \cite{allenPaper}. These beams are qualitatively different from plane waves, being characterized by a helical wave front, and spatial inhomogeneous intensity patterns and momentum distributions \cite{matula2013atomic, peshkov2017photoexcitation,yao2011orbital}. Their unique properties have rendered them a trending topic in the past decade, in particular in atomic physics, where 
 vortex beams are a versatile tool to study  photo-absorption in atomic shell transitions. For example, these beams can be used to address and  separate quantum transitions with multipolarities higher than electric dipole \cite{afanasev2016high,afanasev2018atomic}, to generate rich polarization patterns in microscopic$/$macroscopic samples \cite{afanasev2013off,tw_hydrogenlike,sur_many,peshkov2018rayleigh,schulz2019modification, schmidt2023atomic}, or to suppress the unwanted light shift in atomic clock transitions \cite{schmiegelow2016transfer,lange2022excitation,afanasev2018experimental}.
 A few works have gone further to address theoretically the interaction of vortex beams of higher frequency with nuclei, for instance the manipulation of giant dipole resonances \cite{lu_vortex}, the excitation of multipolar nuclear transitions \cite{kazinski2023excitation}, Delta baryon photo-production \cite{afanasev2022delta} and deuteron photo-disintegration \cite{afanasev2018radiative}.

In this work we investigate theoretically the prospects of using vortex beams to drive the 8 eV nuclear clock transition in $^{229}$Th. This is a unique nuclear transition accessible by vacuum ultraviolet (VUV) light which renders possible a nuclear frequency standard \cite{peik2003nuclear} or a first nuclear laser \cite{tkalya_laser, haowei_laser}. The nuclear clock is a compelling alternative to atomic clocks, potentially with superior accuracy and insensitive to some of the shifts which affect atomic transitions \cite{peik2021nuclear}. As frequency standard based on a nuclear transition, it would also have a significant impact on other fields, for instance the improvement of satellite navigation \cite{thirolf2019improving}, 
the detection of dark matter \cite{peik2021nuclear,Fadeev_2020,tsai2023direct} or investigating temporal variations of fundamental constants \cite{Fadeev_2020, flambaum_2006}.

The 8 eV nuclear transition in $^{229}$Th is a magnetic dipole ($\mathcal{M}1$) and electric quadrupole ($\mathcal{E}2$) multipole mixture, for which beams with orbital angular momentum might be advantageous. Until very recently, the 8 eV isomeric (i.e., metastable) state could only be accessed indirectly, e.g. via $\alpha$-decay of $^{233}$U \cite{Beck2007,von2016direct,Seiferle2019,sikorsky20}, x-ray excitation of the next nuclear excited state at 29 keV \cite{masuda} or $\beta$-decay of $^{229}$Ac \cite{kraemer2023observation}. The first direct VUV laser excitation of the $^{229}$Th isomer was reported just recently \cite{PRL2024}, using plane wave beams. Vortex beams can be generated in optical transmission from regular plane wave beams \cite{Phua2007}, and this technology is scalable to the VUV frequency range. It is therefore timely to address scenarios in which the nuclear clock transition is driven by twisted light beams.

Our study follows the two approaches pursued at the moment to build a nuclear clock \cite{peik2015nuclear}. The first approach 
involves a single-ion nuclear clock in an ion trap \cite{campbell2012single, peik2003nuclear}. This setup is particularly clean and promises great accuracy; however, its realization is technically demanding because of the very low excitation probability per single nucleus. The second approach involves a solid-state nuclear clock using Th-doped VUV transparent crystals \cite{kazakov2012performance, hudson2010, thirolf2024thorium, dessovic2014229thorium,elwell2024laser}. Here, a large number of doped nuclei ($N \approx \num{e14}$ to $\num{e16}$)  can be simultaneously interrogated,  leading to a superior signal to noise ratio and thus a higher stability. The crystal approach leads to nuclear level shifts, splittings and broadenings due to the interactions with the intrinsic electric and magnetic fields of the host crystals \cite{dessovic2014229thorium}. The interaction of the nuclear quadrupole moments with the electric field gradient generated by the crystal lattice leads to the emergence of a nuclear quadrupole splitting on the order \SI{100}{\mega \hertz}.

Our theoretical approach commences with the calculation of the twisted light interaction matrix element, which we  derive for Bessel beams using semi-classical theory and express in terms of the nuclear reduced transition probabilities $B(\mathcal{M}1)$ and $B(\mathcal{E}2)$ and the intensity of the driving field. We then present and discuss nuclear excitation probabilities obtained with the density matrix formalism for single Th ions for both on-axis and off-axis geometry. Compared to plane wave excitation, the interaction with vortex beams leads to a qualitatively different excitation pattern in space. We note here that as it has been shown for atomic systems, the total excitation achieved by vortex beams  averaging over the impact parameter is not higher than with plane wave beams \cite{afanasev2013off,tw_hydrogenlike}. 

For the second nuclear clock approach, we present our results for vortex beams interacting with  an ensemble of nuclei doped in a VUV-transparent crystal. We focus on the quadrupole splitting level scheme of $^{229}$Th doped in a large band gap crystal CaF$_2$ and calculate the nuclear excitation probability induced by a circular polarized twisted light beam and a superposition of two such beams with opposite helicity. The occupation of the nuclear quadrupole levels is represented using an analytical solution of the multilevel Bloch equations. We consider  the cubic symmetry of the crystal and the average over the three possible quantization axes entering the solution of the multilevel Bloch equations. Our simulations show that the vortex excitation in a macroscopic sample can have different features for single quantization axes, but taken into account all crystal orientations, the overall result turns out to be similar to plane wave excitation.

The paper is structured as follows. A short introduction to twisted light and the corresponding derivation of the nuclear interaction matrix element is presented in Sec.~\ref{sec:theory}. Then, our numerical results for the two nuclear clock approaches are presented and discussed in Sec.~\ref{sec:numres}. The paper concludes with a brief discussion in Sec.~\ref{sec:fin}.

\section{Theoretical Background}
\label{sec:theory}
%%%%%%%%%%%%%%%%%%%%%%%%%%%%%%%%%%%%%%%%%%%%%%%%%%5

This section introduces the theoretical background for the interaction between twisted light and nuclei. Though developed independently, our approach resembles the theory work in Ref.~\cite{kazinski2023excitation}.  Throughout this paper, we consider Bessel modes in order to describe the twisted light field. While Bessel-Gauss and Laguerre-Gauss modes are also often used in the literature, 
 Bessel modes are most convenient to be treated analytically and numerically and theoretical predictions have proven so far to 
  match experimental observations quite well \cite{afanasev2018atomic, afanasev2022delta}. For this reason, the section starts with a short introduction to twisted light based on Bessel modes. For a more detailed review on twisted light, we refer to Refs.~\cite{matula2013atomic, peshkov2017photoexcitation}.
We then proceed to address the underlying nuclear interaction matrix element derived in the semiclassical theory, i.e., treating the light field  classically, and the nucleus  quantum mechanically.

\subsection{Bessel beams }
A twisted light beam propagating in the $z$-direction with well defined transverse momentum $\zeta=\ab{k_\perp}$, longitudinal momentum $k_z$, total angular momentum (TAM) projection $m_\gamma$ and helicity $\Lambda$ is characterized by the following vector potential \cite{schulz2020generalized,peshkov_HG}:
\begin{equation}
\label{eq:twVecP}
    \bm{A}^{\text{(tw)}}(\bm{r})= A_0 \int \bm{e}_{\bm{k}\Lambda} e^{i \bm{k} \cdot \bm{r}} a_{\zeta m_\gamma} k_\perp \frac{d k_\perp}{2\pi} \frac{d \alpha_k}{2 \pi}
\end{equation}
with
\begin{equation}
\label{eq:twampl}
    a_{\zeta m_\gamma} = \lrb{-i}^{m_ \gamma} e^{i m_\gamma \alpha_k} \frac{2 \pi}{\zeta} \delta(k_\perp - \zeta)
\end{equation}
being the Fourier amplitude.
Here, $A_0$ is the field amplitude, $\bm{k}$ the wave vector, which is related to the energy of the radiation via $E = \hbar \omega = \hbar c \ab{\bm{k}}$, $\bm{e}_{\bm{k} \Lambda}$ the
polarization vector pointing in the directions determined by $\lrcb{\theta_k, \alpha_k}$ and $\Lambda = \pm 1$ the helicity, respectively.

\begin{figure}
    \centering
    \includegraphics[width = 6 cm]{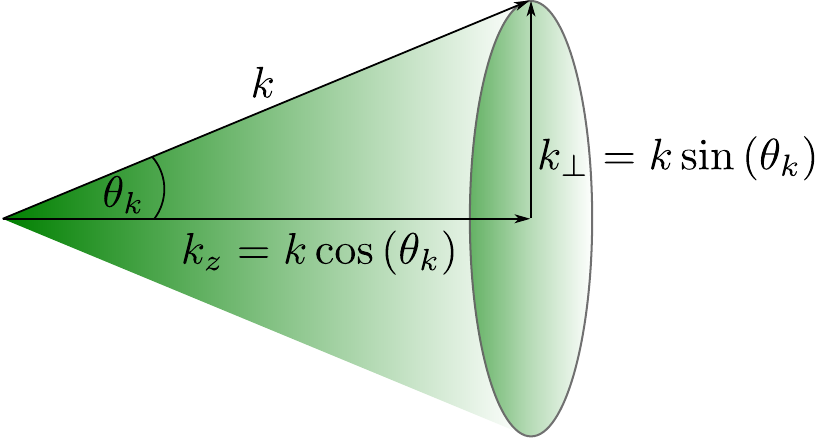}
    \caption{Twisted light beam in k-space. The superposition of the wave vectors $\bm{k}$ form the surface of a cone with constant opening angle $\theta_k = \arctan{(\zeta/k_z)}$. }
    \label{fig:cone}
\end{figure}

In this representation, the Bessel light illustrates a coherent superposition of circularly polarized plane waves where the different $\bm{k}$ vectors span the surface of a cone. This cone has a constant opening angle $\theta_k =\arctan{(\zeta/k_z)}$, the so-called pitch angle, illustrated in Fig.~\ref{fig:cone}.

The real space representation of the vector potential can be determined upon evaluating the integral in Eq.~\eqref{eq:twVecP}. Thereby, in order to integrate over $\alpha_k$, one first needs to expand the polarization vector according to \cite{matula2013atomic}
\begin{equation}
    \bm{e}_{\bm{k} \Lambda} = \sum_{m_s} c_{m_s}e^{-i m_s \alpha_k} \bm{\eta}_{m_s}
\end{equation}
where the sum runs over all possible eigenvalues of the spin angular momentum operator $m_s$.
Here, $\bm{\eta}_{\pm 1} =(1, \pm i,0)/\sqrt{2} $ and $\bm{\eta}_0=(0,0,1)$ are the eigenvectors of the spin angular momentum operator. Moreover, the coefficients $c$ explicitly read
\begin{equation}
   c_{\pm 1}=\frac{1}{2}\lrb{1\pm \Lambda \cos{(\theta_k})}, \, \, \, \, c_0 = \frac{\Lambda}{\sqrt{2}}\sin{(\theta_k)}.
\end{equation}

Rewriting the scalar product 
\begin{equation}
    \bm{k} \cdot \bm{r} = \bm{k}_\perp \cdot \bm{r}_\perp + k_z z = \zeta r_\perp \cos{(\alpha_k -\phi_r)} + k_z z 
\end{equation}
and using the integral representation of the Bessel function \cite{tw_hydrogenlike, schulz2020generalized} 
\begin{equation}
\label{eq:integr-repr-Bessel}
    \int_0^{2\pi} \frac{d\alpha_k}{2\pi} e^{in\alpha_k \pm iz \cos{(\alpha_k)}} = (\pm i)^n J_n(z)
\end{equation}
yields the real space vector potential 
\cite{matula2013atomic, schulz2020generalized}
\begin{equation}
\label{eq:bbreal}
\begin{aligned}
     \bm{A}^{\text{(tw)}}(\bm{r}) &= A_0 e^{ik_z z} \sum_{m_s} i^{-m_s}  c_{m_s}\bm{\eta}_{m_s}\\
    & \times J_{m_\gamma-m_s}(\zeta r_\perp) e^{i(m_\gamma-m_s)\phi_r} .
\end{aligned}
\end{equation}

In the equations above, $J_n$ is the Bessel function of first kind and order $n$. The expression of the real space vector potential  shows that Bessel beams have a spatially inhomogeneous intensity pattern in the transverse plane,  in contrast to plane waves \cite{matula2013atomic}.
Note that usually only paraxial twisted light fields, i.e. $\theta_k<<1$, are available in experiments. As such the sum  \eqref{eq:bbreal} is solely restricted to a single leading term where $m_s =\Lambda$ holds. Furthermore, in the paraxial regime the Bessel beam has a well defined orbital angular momentum projection $m_l$, i.e. the TAM projection decouples to $m_\gamma = m_l + \Lambda$. This means that the beam is not uniquely characterized by $m_\gamma$ but also by certain values for the pair $m_l$ and $\Lambda$.
Apart from that, one can easily show that for $\theta_k = 0$ and $m_\gamma=\Lambda$ the twisted light vector potential recovers the vector potential of a circularly polarized plane wave with an insignificant phase factor $i^{-\Lambda}$.
However, for the sake of generality, we will address also the case of non-paraxial Bessel beams with larger $\theta_k$ values.

\subsection{Nuclear interaction matrix element}
%%%%%%%%%%%%%%%%%%%%%%%%%%%%%%%%%%%%%%%%%%%%%%%%
As a starting point to derive the nuclear interaction matrix element, we consider the semi-classical interaction Hamiltonian for a nucleus in an electromagnetic field given by \cite{AP_superIntense}
\begin{equation}
\label{eq:pwIntHam}
     \mathcal{H}_I = -\frac{1}{c}\int \bm{j}(\bm{r}) \cdot \bm{A}(\bm{r}) d^3\bm{r}.
\end{equation}
Here, $c$ is the speed of light, $\bm{j}$ the nuclear current density and $\bm{A}$ the  vector potential of the impinging electromagnetic field, so far treated as a plane wave. 

In order to take the case of twisted-light-matter interaction into account, we have to implement two modifications in Eq.~\eqref{eq:pwIntHam}. First, we replace the plane wave vector potential by the twisted wave vector potential \eqref{eq:twVecP}. Second, a (two-dimensional) impact parameter $\bm{b}$ is introduced via $e^{-i \bm{k}_\perp \bm{b}}$ in order to specify the position of the nucleus within the spatially inhomogeneous wave front.
Thereby, the impact parameter $\bm{b}=b(\cos{(\phi_b)}, \sin{(\phi_b)},0)$ is defined with respect to the beam center at $r_\perp =0$.
Introducing the expression in Eq.~\eqref{eq:twVecP} in Eq.~\eqref{eq:pwIntHam} and evaluating the integral over $k_\perp$, the interaction Hamiltonian can be written as
\begin{equation}
\label{eq:pwIntHam2}
\begin{aligned}
        \mathcal{H}^{\text{(tw)}}_I &= -\frac{\lrb{-i}^{m_\gamma} A_0}{c} \\
        &\times \int \bm{j}(\bm{r}) \cdot \bm{e}_{\bm{k}\Lambda} e^{i \bm{k} \bm{r}}e^{-i \bm{k}_\perp \bm{b}} e^{i m_\gamma \alpha_k} \frac{d \alpha_k}{2\pi} d^3 \bm{r}.
\end{aligned}
\end{equation}
Note that here the perpendicular component of $\bm{k}$ is now fixed by the Dirac delta function in Eq.~\eqref{eq:twampl}. 
We can further evaluate Eq.~\eqref{eq:pwIntHam2} by expanding $\bm{e}_{\bm{k}\Lambda} e^{i \bm{k} \bm{r}}$ into spherical multipoles according to
\begin{equation}
   \begin{aligned}
        \bm{e}_{\bm{k}\Lambda} e^{i \bm{k} \bm{r}} &=  \sqrt{2\pi} \sum_{L,M}i^L \sqrt{2L+1} \, D_{M \Lambda}^L(\alpha_k, \theta_k,0) \\
        &\times \lrb{\bm{\mathcal{A}}^\mathcal{M}_{LM}+ i \Lambda \bm{\mathcal{A}}^\mathcal{E}_{LM}} .
   \end{aligned}
\end{equation}
Here, $D_{M \Lambda}^L(\alpha_k, \theta_k,0)$ is the Wigner rotation matrix accounting for the different quantization axes of the  light field and of the nucleus.
This matrix can be expanded as $D_{M \Lambda}^L(\alpha_k, \theta_k,0) = e^{-iM\alpha_k}d_{M\Lambda}^L( \theta_k)$ where $d_{M\Lambda}^L( \theta_k)$ is the Wigner small $d$-function.
Moreover, $\bm{\mathcal{A}}^\mathcal{M}_{LM}$ is the magnetic amplitude of multipolarity $L$ and projection $M$, and $\bm{\mathcal{A}}^\mathcal{E}_{LM}$ the electric amplitude, respectively. Using the integral representation of the Bessel function \eqref{eq:integr-repr-Bessel} 
\begin{equation}
   \begin{aligned}
       & \int_0^{2\pi} e^{i(m_\gamma -M)\alpha_k}e^{-i\zeta b \cos{(\alpha_k - \phi_b)}} \, \frac{d\alpha_k}{2\pi}  \\
       &= (-i)^{m_\gamma-M}e^{i(m_\gamma-M)\phi_b} J_{m_\gamma -M}(\zeta b)\, ,
   \end{aligned}
\end{equation}
the interaction Hamiltonian can be written as 
\begin{equation}
\label{eq:hint2}
   \begin{aligned}
        \mathcal{H}^{\text{(tw)}}_I &= -\frac{A_0 \sqrt{2\pi}}{c} \sum_{L,M, \mu} (i\Lambda)^{\delta_{\mu, \mathcal{E}}} i^L (-i)^{2m_\gamma -M}   \\
        &\times e^{i\phi_b (m_\gamma -M)} \sqrt{2L+1}d_{M\Lambda}^L(\theta_k) J_{m_\gamma-M}(\zeta b) \\
        &\times \int \bm{j}(\bm{r}) \cdot \bm{\mathcal{A}}_{LM}^\mu \, d^3 \bm{r}\, ,
   \end{aligned}
\end{equation}
where $\delta_{\mu, \mathcal{E}}$ is the Kronecker delta and $\mu \in \lrcb{\mathcal{E}, \mathcal{M}}$. Again, by setting $\theta_k=0$, the interaction Hamiltonian recovers Eq.~\eqref{eq:pwIntHam} as long as $m_\gamma = \Lambda$.
 The integral in Eq.~\eqref{eq:hint2} can then be evaluated with aid of \cite{AP_superIntense, ring2004nuclear}
 \begin{equation}
   \int \bm{j}(\bm{r}) \cdot \bm{\mathcal{A}}_{LM}^\mu \, d^3 \bm{r} = \frac{k^L c}{i(2L+1)!!}\sqrt{\frac{L+1}{L}}   \mathbb{Q}^\mu_{ L M}
 \end{equation}
where $\mathbb{Q}^\mu_{ L M}$ denotes the multipole moment operator of type $\mu$.
With that a general expression of the Hamiltonian describing the twisted light nucleus interaction is found as
\begin{equation}
 \begin{aligned}
        \mathcal{H}^{\text{(tw)}} &= -\sqrt{2 \pi} E \sum_{L,M, \mu} i^L (-i)^{2m_\gamma -M+1} e^{i\phi_b (m_\gamma -M)}\\
        &\times(i\Lambda)^{\delta_{\mu, \mathcal{E}}}\sqrt{\frac{(2L+1)(L+1)}{L}} \frac{k^{L-1}}{(2L+1)!!} \\
        &\times d_{M \Lambda}^L(\theta_k) J_{m\gamma -M}(\zeta b) \mathbb{Q}^\mu_{ L M}
 \end{aligned}
\end{equation}
where we used $A_0 = E/k$ with $E$ being the amplitude of the electric field. 
Then, the interaction matrix element can then be simply evaluated as
\begin{equation}
\begin{aligned}
    M^{(\text{tw})}_{fi} &= \bra{I_e m_e}    \mathcal{H}^{(\text{tw})}_I \ket{I_g m_g} \\
    &=-\sqrt{\frac{4\pi \mathcal{I}}{c \varepsilon_0}}  \sum_{L,M, \mu} i^L (-i)^{2m_\gamma -M+1} e^{i\phi_b (m_\gamma -M)} \\
    &\times (i\Lambda)^{\delta_{\mu, \mathcal{E}}} \sqrt{\frac{(2L+1)(L+1)}{L}} \frac{k^{L-1}}{(2L+1)!!} \\
        &\times d_{M \Lambda}^L(\theta_k) J_{m\gamma -M}(\zeta b) \bra{I_e m_e} \mathbb{Q}^\mu_{ L M}\ket{I_g m_g}
\end{aligned}
\end{equation}
where $I_{g/e}$ denotes the angular momentum of the nuclear ground$/$excited state and $m_{g/e}$ the respective projections on the nuclear quantization axis. Here, $E= \sqrt{\frac{2\mathcal{I}}{c \varepsilon_0}}$ was expressed in terms of the intensity $\mathcal{I}$ of the driving field where $\varepsilon_0$ denotes the vacuum permittivity.

By applying the Wigner-Eckhart theorem \cite{edmonds1996angular}
\begin{equation}
\begin{aligned}
      &  \bra{I_e m_e} \mathbb{Q}^\mu_{ L M}\ket{I_g m_g} = \frac{(-1)^{I_g - m_g}}{\sqrt{2L+1}} \\
        &\times \braket{I_e m_e I_g -m_g|L M} \langle I_e ||\mathbb{Q}^\mu_{ L M} ||I_g\rangle 
\end{aligned}
\end{equation}
and using the expression of the reduced transition probability \cite{ring2004nuclear}
\begin{equation}
    B(\mu L, I_g \rightarrow I_e)= \frac{1}{2I_g+1} \ab{\langle I_e ||\mathbb{Q}^\mu_{ L M} ||I_g\rangle }^2
\end{equation}
the final form of the twisted nuclear interaction matrix element reads
\begin{equation}
\begin{aligned}
\label{eq:matrixFinal}
        M_{fi}^{(\text{tw})} &= -\sum_{L, \mu} i^L (-i)^{2m_\gamma -\Delta m+1} e^{i\phi_b (m_\gamma -\Delta m)}   \\
        &\times (i\Lambda)^{\delta_{\mu, \mathcal{E}}}  \sqrt{\frac{4 \pi \mathcal{I}}{c \varepsilon_0}}   \sqrt{\frac{(L+1)(2I_g+1)}{L}}   \\ 
        &\times (-1)^{I_g - m_g}    \frac{k^{L-1}}{(2L+1)!!}J_{m\gamma -\Delta m}(\zeta b) d_{\Delta m\Lambda}^L(\theta_k) \\
        &\times \braket{I_e m_e I_g -m_g|L \Delta m}   \sqrt{B(\mu L, I_g \rightarrow I_e)} \\
        &= \sum_{L,\mu} M_{fi,\mu L}^{(\text{tw})}
\end{aligned}
\end{equation}
where the sum over $M$ was dropped due to the selection rule $\Delta m = m_e - m_g$ embodied by the Clebsch-Gordan coefficient.
From this, we can define our Rabi frequency as
\begin{equation}
\label{eq:twR}
    \Omega ^{(\text{tw})}_{fi}= \frac{\ab{M_{fi}^{\text{(tw)}}}}{\hbar}.
\end{equation}

In general, this matrix element describes the interaction of a vortex beam with  any nucleus located at position $b$ with respect to the beam center whose nuclear transitions is described by the reduced transition probability $B(\mu L)$. We observe that for each multipole order $ M_{fi,\mu L}^{(\text{tw})}$, the interaction matrix element factorizes as a product of the standard plane wave matrix element, multiplied by the Bessel function and the Wigner small $d$-function. The latter induces selection rule modifications. A similar factorization has been deduced also in Refs.~\cite{kazinski2023excitation, lu_vortex} and is known also from the interaction of vortex beams with atomic transitions, see, for instance,  Ref.~\cite{afanasev2018atomic}.

\begin{figure}[]
    \centering
    \includegraphics[width = 7cm]{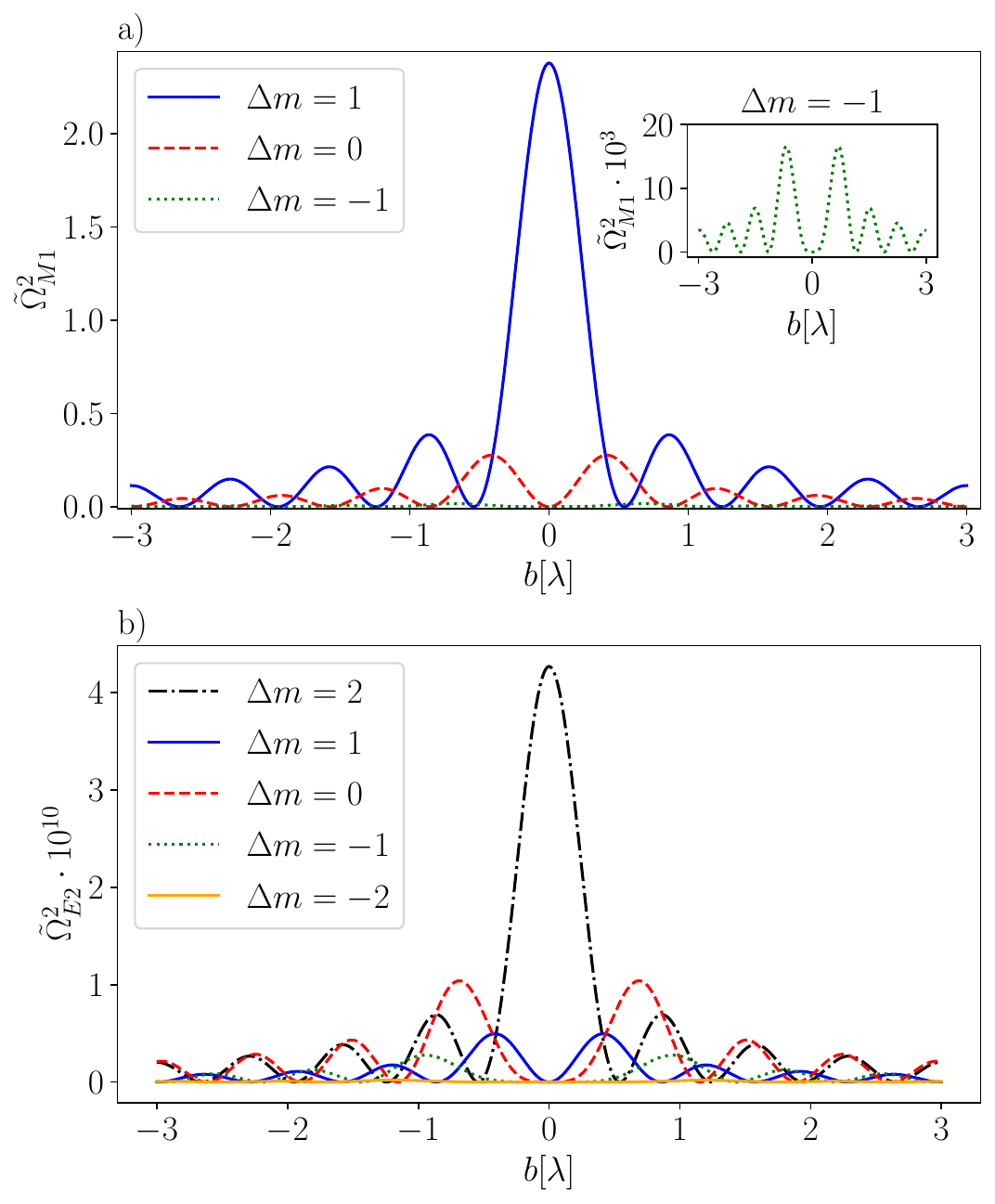}
    \caption{Squared normalized Rabi frequency in units of \si{\metre \squared \per \second \squared \per \watt} as a function of impact parameter $b$ (in units of transition wavelength $\lambda$) for the a) $\mathcal{M}1$ channel and the b) $\mathcal{E}2$ channel of the $^{229}$Th nuclear clock transition. The Rabi frequency is normalized according to $\Tilde{\Omega}^2= \ab{\Omega_{fi,\mu L}^{(\mathrm{tw})}}^2/(\ab{\braket{I_e m_e I_g -m_g|L \Delta m}}^2 \mathcal{I})$. For a), we use $B(\mathcal{M}1, I_g \rightarrow I_e)= 0.017$ W.u. derived with the experimental values in Ref.~\cite{kraemer2023observation}, $m_\gamma = 1$, $\Lambda = 1$ and $\theta_k = \SI{45}{\degree}$. For b), we use $B(\mathcal{E}2, I_g \rightarrow I_e)=27.04$ W.u. taken from the theoretical predictions in Ref.~\cite{minkov2017reduced}, $m_\gamma = 2$, $\Lambda = 1$ and $\theta_k = \SI{45}{\degree}$.}
    \label{fig:matrixElement}
\end{figure}

In the case of $^{229}$Th, the transition from the ground to the excited state proceeds via the $\mathcal{M}1$ channel with a small $\mathcal{E}2$ mixture. Consequently, we can restrict the summation in Eq.~\eqref{eq:matrixFinal} to these two multipole orders. We calculate the square of the Rabi frequency defined in Eq.~\eqref{eq:twR} normalized to the Clebsch-Gordan coefficient of the respective hyperfine transition and to the incoming intensity corresponding to the field amplitude $A_0$. The results are presented in  Fig.~\ref{fig:matrixElement} for each multipole channel of the $^{229}$Th transition individually as a function of impact parameter for different TAM projections $m_\gamma$. The corresponding input parameters used in the calculation are listed in the figure caption.

%%%%%%%%%%%%%%%%%%%%%%%%%%%%%%%%%

The interaction with a twisted light beam leads to the emergence of $2L+1$ transition amplitudes which are position dependent. Due to the spatial modulation by the Bessel function, certain transitions can be enhanced or suppressed relative to each other.
Of particular interest is the impact parameter value $b=0$. In this case, the vortex beam's TAM is completely transferred to the internal degrees of freedom which leads to the selection rule $\Delta m = m_\gamma$. For this reason it becomes possible to address higher order multipole transitions and separate them from lower order contributions. Concretely, for the case of  $^{229}$Th, it is possible to disentangle the $\mathcal{M}1$ + $\mathcal{E}2$ multipole mixing by transferring $\Delta m=2$ units of angular momentum. Then, the transition can only proceed via the $\mathcal{E}2$ channel although the respective radiative rate is almost $10$ orders of magnitude suppressed compared to the $\mathcal{M}1$ channel.

%\textcolor{green}{Maybe remove or rewrite below sentence!}
%Based on these observations, we will make use of the matrix element in the next part and investigate in \ref{single} the temporal dynamics of a single nucleus interacting with a vortex beam for different impact parameters while in part \ref{many} the interaction of a vortex beam with a macroscopic sample where the nuclei are distributed homogeneously over the entire beam is calculated.

\section{Numerical examples}
\label{sec:numres}

In this Section, we present numerical results for nuclear photo-absorption in $^{229}$Th using twisted light. Two experimental configurations based on the two nuclear clock approaches are considered.
First, we investigate the interaction of a single trapped ion with a vortex beam, taking into account both the nuclear hyperfine interaction and Zeeman splitting for the ion level scheme. Thereby, the temporal dynamics induced by resonant driving is investigated for the impact parameter configurations on-axis ($b=0$) and off-axis ($b\neq 0$) for different transitions in the ion. 
Second, we investigate the interaction of twisted light with a macroscopic sample in which the nuclei are distributed homogeneously over the entire whole beam as would be the case of a Th-doped VUV transparent crystal. In particular, we consider the case of the  $^{229}$Th:CaF$_2$ crystal including nuclear quadrupole splitting induced by the crystal fields. We calculate the distribution of the magnetic sublevel population for a circularly polarized Bessel beam and a superposition of two Bessel beams with the same TAM $m_\gamma$, but opposite helicities $\Lambda$.

\subsection{Single nucleus}
\label{single}

%%%%%%%%%%%%%%%%%%%%%%%%%%%%%%%%%%%%%%%%
\begin{figure}
    \centering
    \includegraphics[width = 8cm]{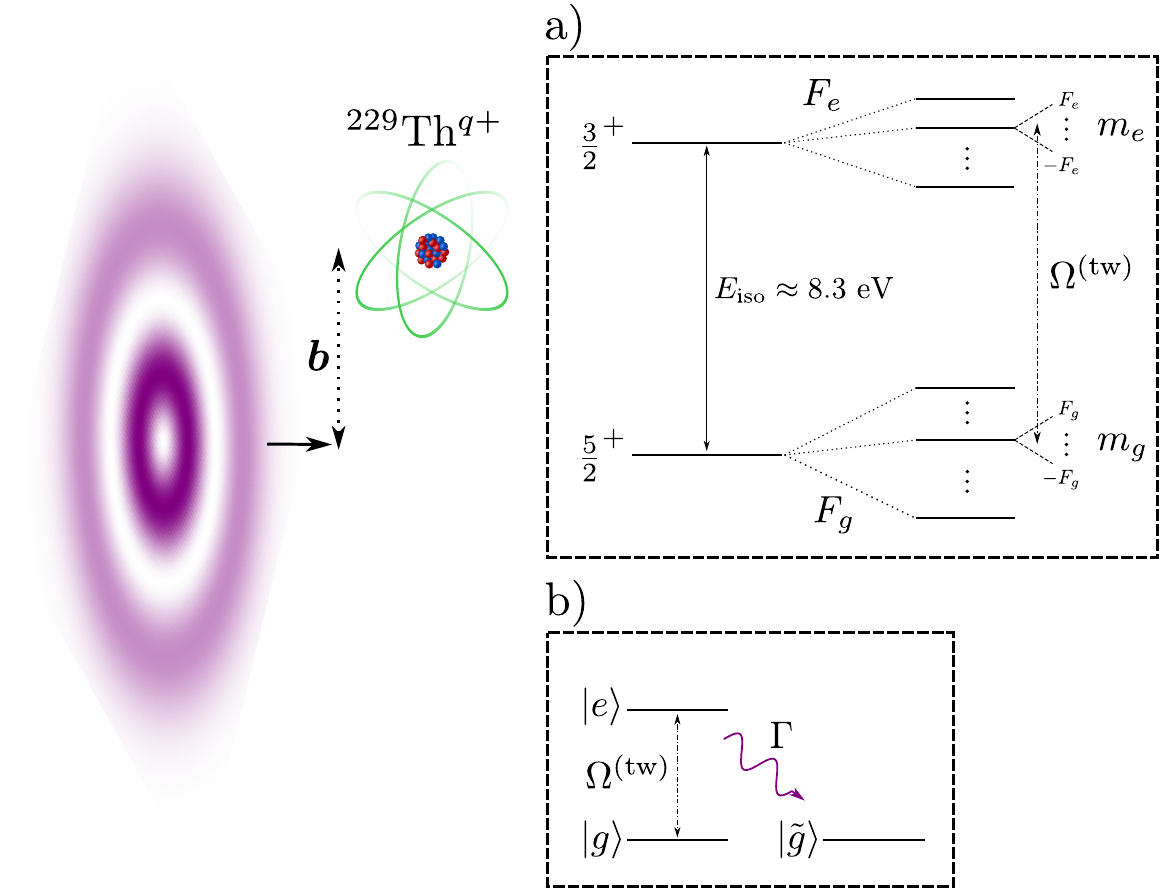}
    \caption{$^{229}$Th$^{q+}$ ion placed at impact parameter $b$ interacting with a resonant vortex beam. a) Level scheme of $^{229}$Th$^{q+}$ in an external magnetic field. b) Effective level system modelling the resonant driving.}
    \label{fig:micro}
\end{figure}
%%%%%%%%%%%%%%%%%%%%%%%%%%%%%%%%%%%%%%%%

The scenario under investigation is illustrated in Fig.~\ref{fig:micro}. A single trapped $^{229}$Th$^{q+}$ ion is placed at impact parameter $b$ in a vortex beam with photon energy resonant to one nuclear transition between the ground and excited state manifolds.  
We consider an external magnetic field  applied in $z$-direction coinciding with the propagation direction of the beam. In the semiclassical approach, the center of mass motion is treated via the well-defined impact parameter. As such, the coupling to rotational degrees of freedom is neglected, see \cite{peshkov2023interaction} for a recent treatment of this effect. The charge state $q$ can in principle be arbitrarily chosen, albeit with some restrictions related to nuclear decay channels that involve the electronic shell and can significantly shorten the nuclear excited state lifetime.  Theoretical calculations have shown that H-, Li- or B-like $^{229}$Th ions display a strong nuclear hyperfine mixing which affects the lifetime of the nuclear excited state \cite{shabaev2022ground}. We do not consider these effects here.

As indicated in panel a) of Fig.~\ref{fig:micro}, the nuclear energy levels experience hyperfine splitting described by the quantum number $F_{g/e}$. The emergence of this splitting is related to the coupling of nuclear and electronic degrees of freedom. Here, the quantum number $F_{g/e}$ runs from $\ab{J-I_{g/e}}\leq F_{g/e}\leq J +I_{g/e}$ where $J$ is the angular momentum quantum number of the electronic shell and $I_{g/e}$ the angular momentum quantum number of the nuclear ground$/$excited state.
In order to specify the quantization axis of the system, an external weak magnetic field is applied. This leads to a further splitting of the energy levels $F_{e/g}$ where each level splits into $2F_{e/g}+1$ Zeeman sublevels running from $-F_{e/g}\leq m_{e/g} \leq  F_{e/g}$.
We will consider resonant driving  with a twisted light field for different Zeeman $\Delta m = m_e -m_g$ transitions.

The multilevel system is effectively modelled as a three level system described by the density matrix $\rho= \sum_{e,g,\Tilde{g}}\rho_{ij}$ where the diagonal elements correspond to the level populations and the off-diagonal elements to the coherences, respectively. Here, the resonant driving occurs between a fixed ground $\ket{g}$ and excited $\ket{e}$ state. We model the relaxation processes proceeding via a quasi degenerate ground state $\ket{\Tilde{g}}$ which accounts for the nuclear excited state population decaying to a variety of other magnetic or hyperfine substates.

%The experimental scenario is illustrated in Fig.~\ref{fig:micro}. A single $^{229}$Th$^{3+}$ ion  or a small ensemble is localized in a trap. An external magnetic field is applied in z-direction coinciding with the propagation direction of the wave. Here, the electronic shell remains in its $5F_{5/2}$ ground state. Due to the coupling of nuclear and electronic degrees of freedom a hyperfine splitting emerges. This leads to the formation of $2I +1$ nuclear substates labelled as $F$. In combination with the external magnetic field the hyperfine levels further split into $2F+1$ Zeemann sublevels which will be used for the quantum transitions.

%Thereby, resonant driving is considered between the $\ket{5F_{5/2}, I_g = 5/2; F_g=5, m_g} \leftrightarrow \ket{5F_{5/2}, I_e = 3/2; F_e=5, m_e}$ manifold for different $\Delta m = m_e -m_g$ transitions.
%\textcolor{red}{Note: This publication does not aim at performance issues of a nuclear clock.}

%The nucleus is effectively modelled as a three level system embodied by the density matrix $\rho$. Here, the resonant driving occurs between a fixed ground and excited state while spontaneous emission occurs to a quasi degenerate ground state $\ket{\Tilde{g}}$. This is related to the fact that the nuclear excited state population can decay to variety of other magnetic substates.

The dynamics of the system are governed by the Master equation \cite{scully1999quantum}
\begin{equation}
\label{eq:eom}
    \dot{\rho} = \frac{1}{i\hbar} \lrsb{\mathcal{H}, \rho} + \mathcal{L}[\rho]
\end{equation}
where $\mathcal{H}$ is the Hamiltonian describing the light-matter interaction and $\mathcal{L}[\rho]$ the Lindblad operator accounting for relaxation processes, respectively.
The concrete form of the Hamiltonian (on resonance) is given by
\begin{equation}
    \mathcal{H} = - \hbar \lrb{\Omega_{eg}^{(\text{tw})}\ket{e}\bra{g} + \, \text{h.c.} }
\end{equation}
where $\Omega_{eg}^{(\text{tw})}$ is the twisted light Rabi frequency driving the transition $\ket{g}\rightarrow \ket{e}$ whose general expression is given in Eq.~\eqref{eq:twR}. For a proper description of the system including hyperfine splitting, the Rabi frequency must be  modified  by replacing the Clebsch-Gordan coefficient in Eq.~\eqref{eq:matrixFinal} with \cite{von2020theory}
\begin{equation}
\label{eq:coeff}
 \begin{aligned}
      &  \braket{I_e m_e I_g -m_g|L \Delta m} \rightarrow \sqrt{2F_e +1}\sqrt{2F_g+1}\\
       &\times  \braket{F_e m_e F_g -m_g|L \Delta m} \begin{Bmatrix}
    I_g & L & I_e \\
    F_e & J & F_g
\end{Bmatrix}\, ,
 \end{aligned}
\end{equation}
 where $ \{...\}$ denotes the Wigner-$6J$ symbol. Note that this transformation only holds if the electronic shell remains in its state $J$ during the radiative coupling.

Moreover, the Lindblad operator reads
\begin{equation}
\label{eq:LB}
    \begin{aligned}
        \mathcal{L}[\rho] &= \Gamma \rho_{ee}\lrb{\ket{\Tilde{g}}\bra{\Tilde{g}}-\ket{e}\bra{e}}\\
        &-\lrb{\frac{\Gamma +\Gamma_\ell}{2}} \lrb{\rho_{eg}\ket{e}\bra{g}+ \, \mathrm{h.c.}},
    \end{aligned}
 \end{equation}
where we have neglected relaxation between sublevels of the same manifold and relaxation of the coherences involving $\ket{\Tilde{g}}$, since these are not affected by resonant driving. The total relaxation rate $\Gamma$ is thereby given by $\Gamma = \Gamma_\gamma + \Gamma_{\text{el}}$ with $\Gamma_\gamma$  the radiative decay rate and $\Gamma_{\text{el}}$ the nuclear decay rate via energy transfer to the electronic shell, respectively. The nucleus can transfer its excitation energy to the electronic shell via internal conversion or electronic bridge (EB) processes. Internal conversion refers to the process in which the nuclear excited state decays by transferring its energy to an shell electron which is then expelled into the continuum. Since the second ionization potential of Th lies at 12 eV, this channel should be energetically forbidden already for the first ionized state $^{229}$Th$^+$. In turn, EB describes the energy transfer between nucleus and electronic shell without a change in the ionic charge and with an accompanying photon emission to account for the energy mismatch of available electronic and nuclear states \cite{porsev+,porsev3+}.

Moreover, $\Gamma_\ell$ in Eq.~\eqref{eq:LB} denotes the laser bandwidth which leads to a faster dissipation of the coherences.
For the radiative decay rate we take $\Gamma_\gamma = \SI{3.14e-4}{\per \second}$ \cite{kraemer2023observation}, while the value $\Gamma_{\text{el}}$ depends on the ion species under investigation.
Upon solving Eq.~\eqref{eq:eom} numerically with the initial condition $\rho_{gg}(0)=1$, we can determine the temporal dynamics of the nuclear excited state population as a function of impact parameter.
For all upcoming calculations, we consider a vortex beam with TAM projection $m_\gamma = 2$ and $\Lambda = 1$ interacting with a well localized  $^{229}$Th$^{q+}$ target for different opening angles.

\subsubsection{On-axis ($b=0$)}
%%%%%%%%%%%%%%%%%%%%%%%%%%%%%%%%%%%%%
For zero impact parameter, the selection rule  $\Delta m = m_\gamma$ comes into play. Thus, by choosing $\Delta m=2$, only the $\mathcal{E}2$ channel can be driven on-axis, such that the two multipolarities of the nuclear clock transition can be completely spatially separated. For the isomeric transition in $^{229}$Th, the radiative rate for the $\mathcal{E}2$ channel is almost 10 orders of magnitude suppressed compared to the $\mathcal{M}1$ one. This means that while in principle one could drive pure  $\mathcal{E}2$ Rabi oscillations, this requires either a very large Rabi frequency or a moderate Rabi frequency and extremely long interrogation times. 
In both cases, ion loss from the trap due to multi-photon ionization or chemical reactions with background gases may occur. 
Moreover, a long interrogation time is limited by the relaxation of the excited state population.
A way to reduce unwanted ionization is to choose a higher ion charge state for this case, for instance $^{229}$Th$^{35+}$, discussed in the context of  electron beam ion trap generation via electron-atom collisions in Ref.~\cite{Pavlo_prl}. We consider in the following this charge state with the electron shell in its $J=15/2$ ground state. Based on the analysis in Ref.~\cite{Pavlo_prl} and the lack of electronic states with energies matching the nuclear transition energy, we neglect the spontaneous EB decay rate in the calculation and set $\Gamma_{\text{el}}=0$.

For nuclear excitation, we envisage the $\ket{F_g = 5 \, m_g = 5}\rightarrow \ket{F_e = 7 \, m_e =7}$ transition, since the corresponding product of Clebsch-Gordan coefficient and Wigner $6$-$J$ symbol in Eq.~\eqref{eq:coeff} is largest. 
Although the transition is per se of $\mathcal{E}2$ multipolarity, due to the selection rule $\Delta F =2$, resonant driving is optimized with the choice of $\Delta m = 2$. We note that for the on-axis case ($b=0$), regardless of the  choice of $\Delta F$, only  the    $\mathcal{E}2$ transition will be driven once we choose $m_\gamma=2$. 

%%%%%%%%%%%%%%%%%%%%%%%%%%%%%%%%%%%%%
\begin{figure}
    \centering
    \includegraphics[width = 8cm]{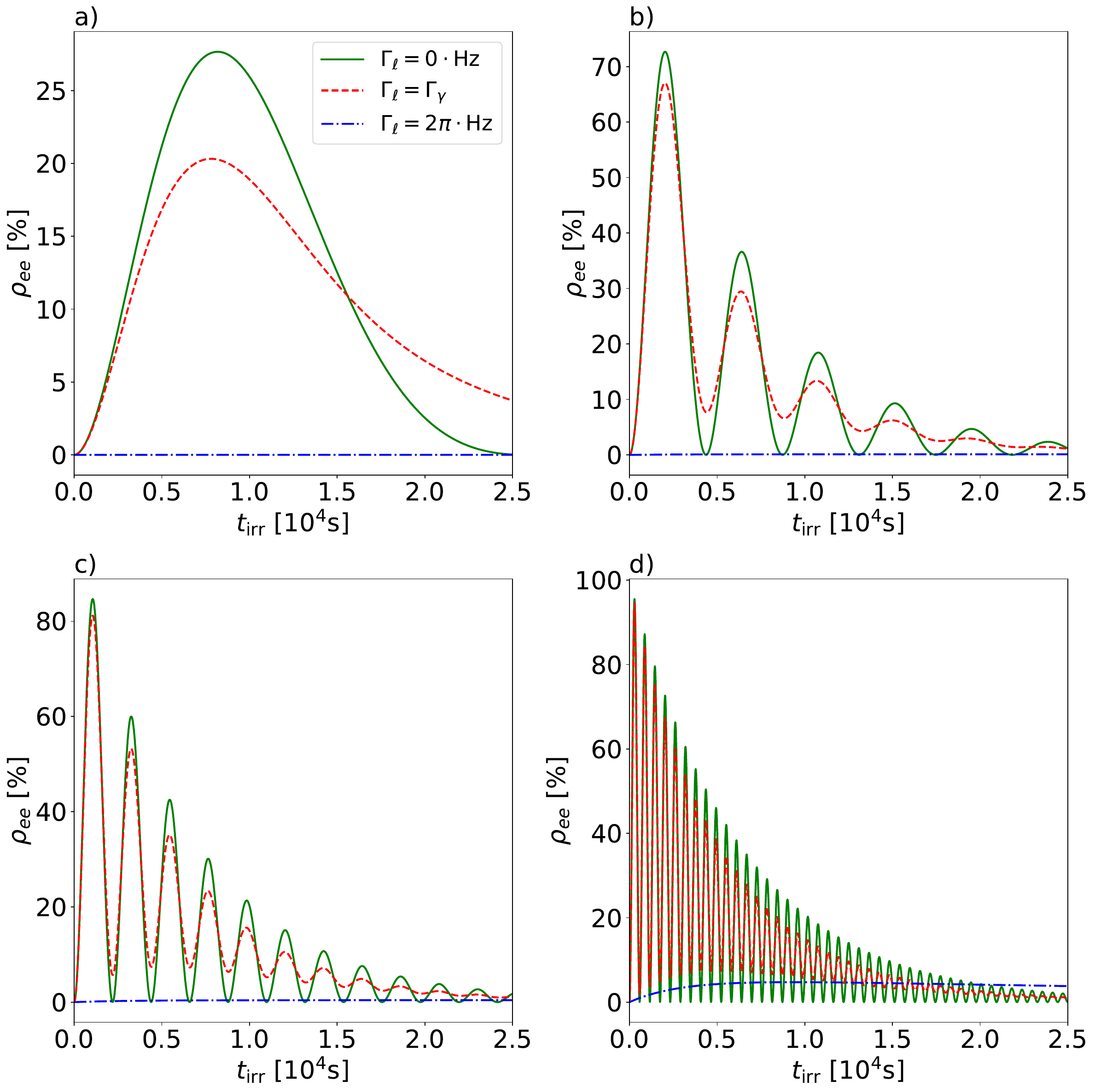}
    \caption{Calculated excited state population as a function of irradiation time for the      
    resonant $\mathcal{E}2$ driving considering a single $^{229}$Th$^{35+}$ ion in the center of a vortex beam  with $m_\gamma = 2$ and $\Lambda = +1$ for different laser bandwidths. The pitch angles of the beams are  a) $\theta_k = \SI{1}{\degree}$,  b) $\theta_k= \SI{5}{\degree}$, c) $\theta_k = \SI{10}{\degree}$ and d) $\theta_k = \SI{60}{\degree}$.}
    \label{fig:e2}
\end{figure}
%%%%%%%%%%%%%%%%%%%%%%%%%%%%%%%%%%%%

In Fig.~\ref{fig:e2} we present our numerical results for resonant excitation considering  different opening angles and laser bandwidths.  
We choose a fixed set of ground and excited nuclear states and consider a fictitious laser intensity value of   $\mathcal{I}=\SI{15}{\watt \per \centi \metre \squared}$. The excited state population as a function of the irradiation time for small $\theta_k$ angles, also known as the paraxial regime, is presented in Fig.~\ref{fig:e2}a). For a perfectly coherent cw laser $(\Gamma_\ell = 0 \cdot\si{\hertz})$, the excitation probability reaches a maximum of only $\approx\SI{27}{\percent}$ and drops down back to zero completing a Rabi cycle in a few hours of interrogation time.  The situation is similar for a laser with a bandwidth of the order of the radiative decay rate $(\Gamma_\ell = \Gamma_\gamma)$
 with the main difference that the amplitude is slightly suppressed. For a more realistic $\Gamma_\ell = 2\pi \cdot\si{\hertz}$ corresponding to the bandwidth of a mode of narrowband frequency comb with high repetition rate \cite{von2020theory, von2020concepts}, no excitation via the $\mathcal{E}2$ channel is achieved.

The driving becomes more efficient for  larger pitch angles. We consider the vicinity of the paraxial regime in Figs.~\ref{fig:e2}b)-c). Several Rabi cycles with a maximal excitation probability of $\approx \SI{70}{\percent}$ to $\SI{84}{\percent}$ can be passed within a few hours time span for $\Gamma_\ell = 0\cdot\si{\hertz}$ and $\Gamma_\ell = \Gamma_\gamma$. Compared to the paraxial regime results, this is due to the larger Rabi frequency which results from the Wigner small  $d$-function  
\begin{equation}
    d_{21}^2 (\theta_k) = 2 \cos{\left(\frac{\theta_k}{2}\right)}^3\sin{\left(\frac{\theta_k}{2}\right)}
\end{equation}
in Eq.~\eqref{eq:matrixFinal}. The  Wigner small $d$-function is increasing in the angular interval $[0^\circ,60^\circ]$, such that  a higher pitch angle increases the driving efficiency. We therefore present in Fig.~\ref{fig:e2}d) the excitation probability for 
$\theta_k=60^\circ$. In this case, several Rabi cycles can be passed for $\Gamma_\ell = 0\cdot\si{\hertz}$ and $\Gamma_\ell = \Gamma_\gamma$ already within one hour with excitation probabilities approaching a maximum probability of $\approx$ \SI{95}{\percent}. Only at this pitch angle can also a narrowband frequency comb  produce a non-negligible excitation probability, albeit only the small value of  $\approx$ \SI{5}{\percent} reached after approx.~2.5 hours.

Overall, pure Rabi oscillations of the $\mathcal{E}2$ channel alone seem to be very challenging experimentally. We note that the nuclear reduced transition probability $B(\mathcal{E}2)$ has not been determined experimentally so far. Thus, the prospect of an experiment which would lead to its measured value is appealing. However, practical considerations show that unrealistically large intensities and long interrogation times would be required. In addition, one requires an accurate confinement of the particle at zero impact parameter over a long period of time. Small deviation from $b=0$ might open the $\mathcal{M}1$ channel which would immediately dominate the excitation process due to its much stronger nuclear reduced transition probability. We thus conclude that the on-axis case with its challenges is of rather academic interest and turn to the next case of off-axis driving, i.e., at non-zero impact parameter.

\subsubsection{Off-axis ($b\neq 0$)}
Once off-axis, the special selection rule $\Delta m = m_\gamma$ no longer acts and we will have $\mathcal{M}1$ + $\mathcal{E}2$ multipole mixing. Due to the much stronger $\mathcal{M}1$ radiative channel, we will neglect for this case the $\mathcal{E}2$ multipole mixing. We therefore consider transitions with $\Delta m \in \lrcb{-1,0,1}$. The required laser intensities for achieving a reasonable fraction of excited nuclei is less dramatic and the corresponding interrogation time shorter. Therefore, one can envisage also lower ionic charge states. 
In the following, we consider $^{229}$Th$^{3+}$ which is the most promising ion species for a single ion nuclear clock \cite{campbell2012single, peik2003nuclear}. Similar to assumptions made in the on-axis case, the electronic shell of $^{229}$Th$^{3+}$ remains in its $J=5/2$ ground state during the radiative driving. We consider three $\mathcal{M}1$ transitions between sublevels of the ground and excited nuclear levels of $^{229}$Th$^{3+}$ with hyperfine splitting 
\begin{equation}
    \begin{aligned}
        &\ket{F_g=5 \, m_g =\pm5} \rightarrow  \ket{F_e=4 \, m_e =\pm4} \, ,\\
        &\ket{F_g= 5\, m_g =0} \rightarrow \ket{F_e=4 \, m_e =0}\, ,
    \end{aligned}
\end{equation}
where the former two were proposed as clock transitions \cite{ campbell2012single}.
Since the electronic shell remains in its ground state, the spontaneous EB decay is on the order of $\Gamma_{\mathrm{eb}} \approx \mathcal{O}(\SI{e-5}{\per \second})$ \cite{porsev3+}, about one order of magnitude smaller than the radiative decay rate. We therefore consider $\Gamma \approx \Gamma_\gamma$.

%%%%%%%%%%%%%%%%%%%%%%%%%%%%%%%%%%%%%%%%%%%%
\begin{figure*}[]
    \centering
    \includegraphics[width = 17cm]{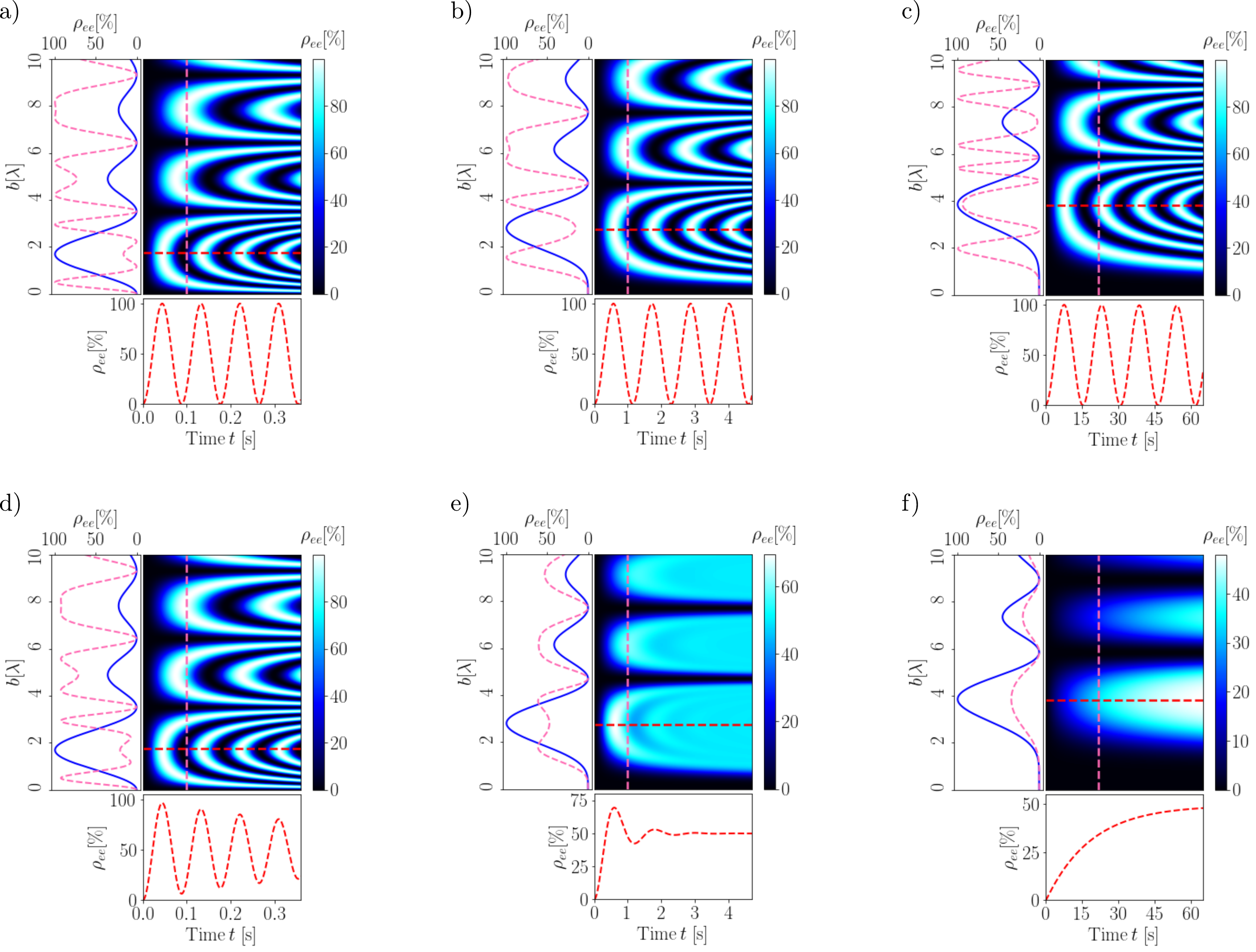}
    \caption{Population dynamics as a function of impact parameter $b$ (in units of the transition wave length $\lambda$) and time $t$ for three $\mathcal{M}1$ transitions in a $^{229}$Th$^{3+}$ ion. Here, $m_\gamma = 2$, $\Lambda = 1$, $\theta_k=\SI{10}{\degree}$ and $ \mathcal{I}=\SI{0.5}{\watt \per \centi  \metre \squared}$ for all calculations. a) $\ket{F_g =5 \,m_g = -5}\rightarrow \ket{F_e =4\, m_e = -4}$ for $\Delta m = 1$ and $\Gamma_\ell = 0\cdot  \si{\hertz}$. b) $\ket{F_g =5 \, m_g = 0}\rightarrow \ket{F_e =4 \, m_e = 0}$ for $\Delta m = 0$ and $\Gamma_\ell = 0\cdot  \si{\hertz}$. c) $\ket{F_g =5 \,  m_g = 5}\rightarrow \ket{F_e =4 \, m_e = 4}$ for $\Delta m = -1$ and $\Gamma_\ell =0\cdot  \si{\hertz}$. d)-f) same as  a)-c), with the difference $\Gamma_\ell = 2\pi \cdot \si{\hertz}$.}
    \label{fig:vortexTable}
\end{figure*}
%%%%%%%%%%%%%%%%%%%%%%%%%%%%%%%%%%%%%%%%%%%%%

We calculate the excited state population $\rho_{ee}$ for nuclear transitions  driven resonantly with a tuneable vortex beam with parameters $\theta_k = \SI{10}{\degree}$, $m_\gamma=2$, $\Lambda=1$ and $\mathcal{I}=\SI{.25}{\watt \per \centi \metre \squared}$.
The population dynamics of the excited state $\rho_{ee}$ as a function of position and time for two laser band widths $\Gamma_\ell = 0 \cdot \si{\hertz}$ (cw) and $\Gamma_\ell = 2\pi \cdot \si{\hertz}$ are presented in Fig.~\ref{fig:vortexTable}. The case of a completely coherent cw laser ($\Gamma_\ell=0\cdot  \si{\hertz}$) is presented in the colormaps in Figs.~\ref{fig:vortexTable}a) for $\Delta m =1$, b) for $\Delta m =0$ and c) for $\Delta m =-1$, respectively. The excited state population is presented as a function of impact parameter and time. 
Bright regions in the map display high excitation probabilities while dark regions display low probabilities. In Fig.~\ref{fig:vortexTable}a) we observe that strong driving of the $\mathcal{M}1$ transition for $\Delta m = 1$ occurs already at small impact parameters, only slightly off-axis.  Since the ion impact parameter cannot be kept precisely fixed in the trap, this result supports our conclusion of the previous subsection that  pure resonant $\mathcal{E}2$ driving at $b=0$ is rather unrealistic with current experimental capabilities.

Along the abscissa, the excited state population exhibits Rabi oscillations resulting from resonant driving. Along the ordinate, $\rho_{ee}$ 
is modulated according to $J_{m_\gamma-\Delta m}^2(\zeta b)$. For a fixed time indicated by the vertical dashed line, we  present on the panels left of the colormaps the corresponding projected impact-parameter dependence of the excited state population. For a better understanding of the displayed minima, we plot in each projection also the corresponding Bessel function squared (normalized to unity) to distinguish between the zeros of $J_{m_\gamma-\Delta m}^2(\zeta b)$ and Rabi oscillation minima due to the chosen time point. 

For a fixed impact parameter value (indicated by the horizontal dashed line), we present projections in the panels below the colormaps to show the excited state population as a function of time. Since the Wigner small-$d$ functions entering the expression of the Rabi frequency obey the relation
\begin{equation}
\label{eq:wignerrel}
d_{11}^1(\theta_k)>d_{01}^1(\theta_k)>d_{-11}^1(\theta_k)
\end{equation}
for the considered pitch angle $\theta_k = \SI{10}{\degree}$, the transition $\ket{F_g=5 \, m_g =-5} \rightarrow  \ket{F_e=4 \, m_e =-4}$ with $\Delta m=1$ is driven on the fastest timescale. We note that the order of the Wigner functions in Eq.~\eqref{eq:wignerrel} is reversed if we choose $\Lambda=-1$. Correspondingly, the timescales of the Rabi cycles for the three $\Delta m=0,\pm1$ cases is also reversed for $\Lambda=-1$: for the same pitch angle,  the transition $\Delta m = -1$ is fastest while $\Delta m = 1$ is slowest. The times required for a Rabi cycle however follow  $J_{m_\gamma-\Delta m}^2(\zeta b)$ and will therefore not have the same values as for the case $\Lambda=1$.
We note that tuning the pitch angle can also manipulate the duration of a Rabi cycle.  However, by means of varying $\theta_k$ also the vertical modulation via $J_{m_\gamma-\Delta m}^2(\zeta b)$ is influenced such that the zeros of the Bessel function shift.

In case of $\Gamma_\ell = 2 \pi \cdot \si{\hertz}$, the oscillation patterns of the excited state population $\rho_{ee}$ become more washed out. Although the vertical Bessel modulation in impact parameter is conserved, the temporal dynamics in terms of Rabi oscillations differ. 
The smallest modifications occur in the case of  $\Delta m = 1$, where the Rabi oscillations present a slightly decaying envelope. The dynamics for $\Delta m = 0$ and $\Delta m = -1$ are mediated by incoherent pumping resulting from a faster relaxation of the coherences in Eq.~\eqref{eq:LB}. This leads to the suppression or even absence of Rabi oscillations together with a reduced amplitude as it is shown in the projection of the colormaps. However, coherent dynamics can be recovered by either increasing the pitch angle or the intensity provided that the condition $\Omega>> \Gamma_\ell/2$ is fulfilled \cite{von2020theory}.

The off-axis setup can be used to probe different $\mathcal{M}1$ transitions in $^{229}$Th$^{3+}$ with a single tunable laser beam. The selection rules for vortex beam excitation allows driving of all three transitions; however, a narrow-band tunable laser will selectively drive just the transition to which its frequency is resonant. This can be interesting for metrology applications, for instance comparing clock transitions.
The excitation strength can be tuned with respect to the position of the nucleus in the wave front according to the projection in Fig.~\ref{fig:vortexTable} or by varying the pitch angle.   We note that in
 comparison with the on-axis case discussed in the previous Subsection, this scenario seems to be more feasible, since shorter interrogation times and lower laser intensities are required.
According to Refs.~\cite{von2020concepts,peik2021nuclear,zhang2022tunable} a high repetition frequency comb with comb mode bandwidth $\Gamma_\ell= 2 \pi \cdot \si{\hertz}$ and $\mathcal{I}=\mathcal{O}( \SI{1}{\watt \per \centi \metre \squared})$ should be available in the near future.

The Bessel modes used here represent an idealization, similar to the frequently used idealized plane wave picture. Modelling a real physical scenario requires the usage of Laguerre-Gaussian modes. However, close to the beam center and in the paraxial approximation we expect Laguerre-Gauss and  Bessel modes to predict a similar behaviour \cite{afanasev2018experimental}. When considering 
Laguerre-Gaussian beams, we expect a temporally finite modulation with slightly modified spatial pattern of the excitation. This is however beyond the scope of the present work.

\subsection{Ensembles of nuclei}
\label{many}
%%%%%%%%%%%%%%%%%%%%%%%%%%%%%%%%%%%%%%%
In the second part of this Section, we consider the interaction of twisted light with  a macroscopic $^{229}$Th target where the nuclei are distributed homogeneously over the entire beam.
Here, our model system is $^{229}$Th:CaF$_2$. The host crystal CaF$_2$ is an ideal environment for doping $^{229}$Th. The band gap of $\approx$ $\SI{11}{\electronvolt}$-$\SI{12}{\electronvolt}$ \cite{rubloff1972far,barth1990dielectric,tsujibayashi2002spectral} renders the crystal transparent at the wavelength of the nuclear clock transition. In addition, high doping densities upwards $n\approx \SI{e17}{\per \centi \metre \cubed}$ can be achieved \cite{beeks2023optical}. As a consequence, a larger number of nuclei can be interrogated at the same time leading to an improved clock stability. 

\begin{figure}
    \centering
    \includegraphics[width = 7.5cm]{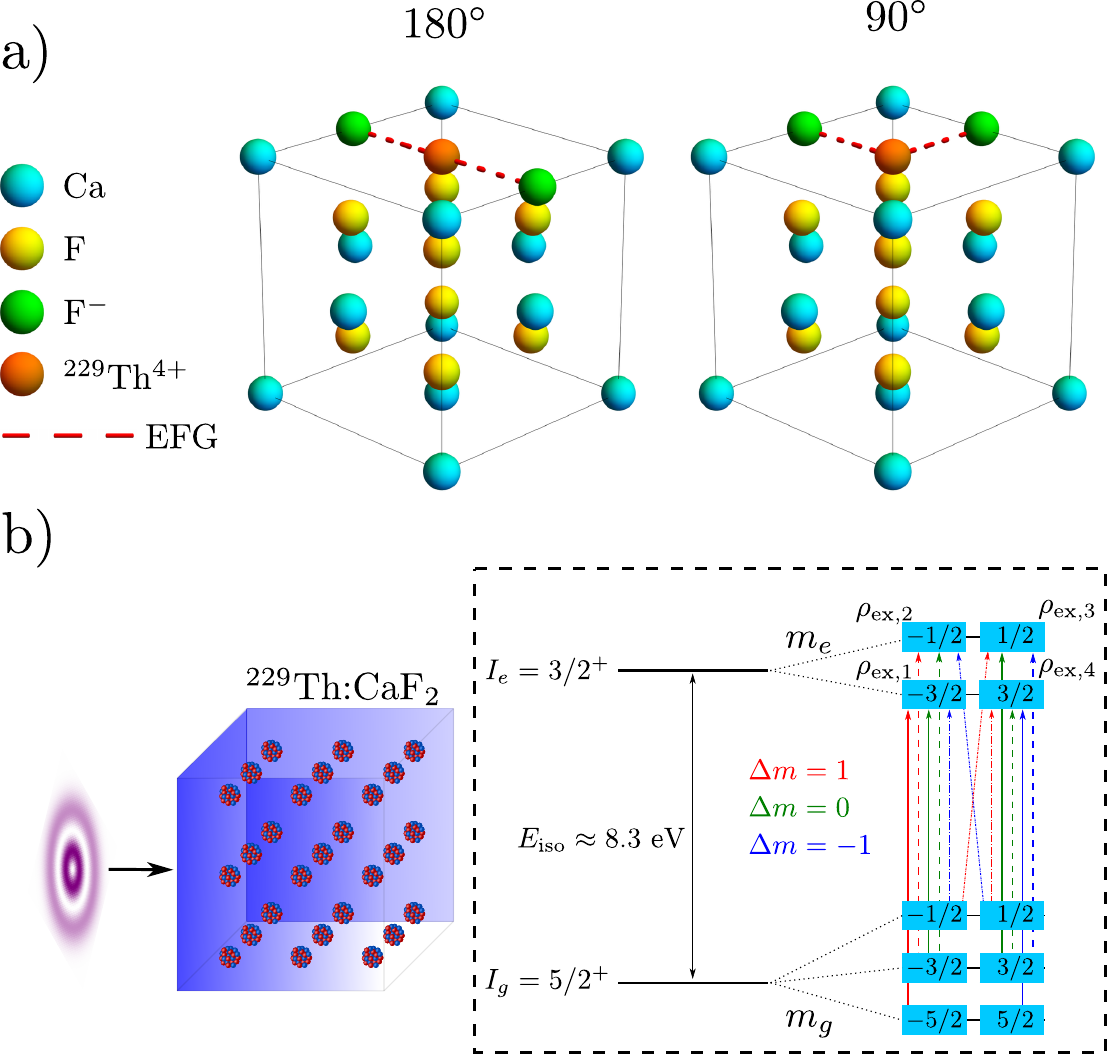}
    \caption{ a)$^{229}$Th:CaF$_2$ crystal structure with preferred charge compensation directions. Adapted from Ref.~\cite{nickerson2018collective}.
    b)$^{229}$Th:CaF$_2$ crystal interacting with twisted light. The nucleus experiences electric quadrupole hyperfine splitting due to the coupling to the electronic surrounding. This leads to the emergence of $6$ ground states and $4$ excited states. A broadband pulse, much broader than the splitting of a hyperfine level $\Gamma_\ell >> \mathcal{O}(\si{\mega \hertz})$, drives all 12 available $\mathcal{M}1$ transitions simultaneously.}
    \label{fig:manyScenario}
\end{figure}

The underlying cubic lattice hosts $^{229}$Th in a charge state $4+$ leading to a suppression of the internal conversion decay channel. Within the lattice, one of the calcium ions is replaced by the thorium ion and two more fluorine interstitial ions emerge for charge compensation. Ab initio DFT studies predict several preferred doping configurations \cite{dessovic2014229thorium}, where two are of particular interest \cite{nickerson2018collective}. The fluorine ions are either in a \SI{90}{\degree} or \SI{180}{\degree} configuration as illustrated in Fig.~\ref{fig:manyScenario}a). In contrast to the free ionic system, thorium doped CaF$_2$ experiences a quadrupole splitting according to \cite{collins1967electric}
\begin{equation}
    \hat{\mathcal{H}}_{\mathcal{E}2} = \frac{eQ V_{zz}}{4I(2I-1)}\lrsb{3 \hat{I}_z- \hat{I}+\frac{\eta}{2}(\hat{I}_+^2+\hat{I}_-^2)}.
\end{equation}
Here, $e$ denotes the electric charge, $Q$ are the nuclear quadrupole moments, $V_{zz}$ is the dominant component of the electric field gradient (EFG) at the thorium nucleus and $\eta = (V_{xx}-V_{yy})/V_{zz}$ is the asymmetry parameter of the EFG. The operators $\hat{I}$ and $\hat{I}_z$ denote the angular momentum and angular momentum projection operators, while $\hat{I}_{+/-}$ are the raising$/$lowering operators. For the sake of simplicity, we consider throughout this section the case of the \SI{180}{\degree} configuration where $\eta =0$.
In that case, the nuclear level scheme splits into $(2I_g+1)$ ground and $(2I_e+1)$ excited states of different energies as shown in Fig.~\ref{fig:manyScenario}b).
Furthermore, for this configuration the quantization axis is not determined by an external magnetic field but instead by the orientation of the EFG along the F$^-$-$^{229}$Th$^{4+}$-F$^-$ bond.

%%%%%%%%%%%%%%%%%%%%%%%%%%
%%%%%%%%%%%%%%%%%%%%%%%%%%%%

For monitoring the excitation of the isomeric state
in the crystal without relying on fluorescence, radiofrequency nuclear quadrupole resonance spectroscopy
(NQRS) of the excited state hyperfine splitting manifold has been proposed \cite{peik2021nuclear}.   To detect a NQRS signal, a certain population difference, i.e., a polarization, between the involved
levels needs to be implemented. NQRS cannot discern
between sublevels with the same absolute value of the
spin projection $|m_e|$. Thus, the polarization of excited
state populations should occur between $\rho_{\mathrm{ex},1}+\rho_{\mathrm{ex},4}$ and  
$\rho_{\mathrm{ex},2}+\rho_{\mathrm{ex},3}$.  This seems extremely challenging thermally,
so in the following we investigate how the interaction with
vortex beams can produce such a polarization.

\begin{figure*}
    \centering
    \includegraphics[width =14cm]{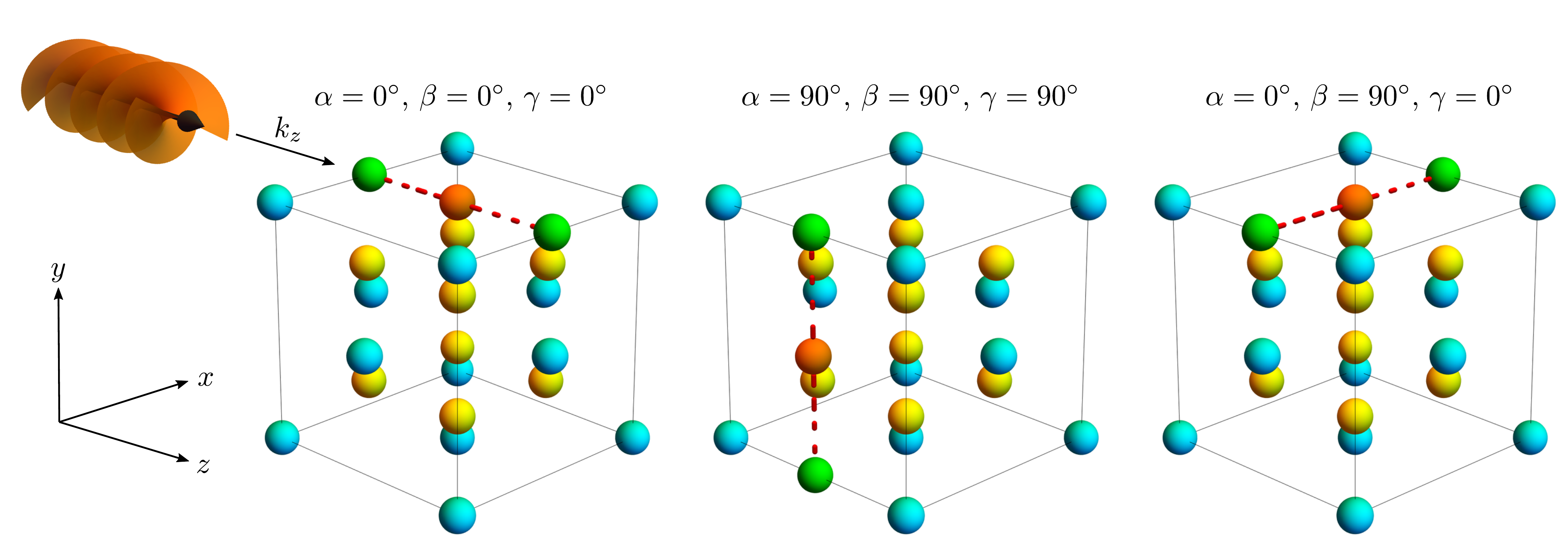}
    \caption{$^{229}$Th:CaF$_2$ crystal structure with $\SI{180}{\degree}$ fluoride interstitials. Here, all three possible orientations of the EFG and the corresponding Euler angles  with respect to the laboratory reference frame (left)  are illustrated.}
    \label{fig:orientations}
\end{figure*}

For our calculations, we assume that a weak, broadband vortex field with arbitrary TAM projection $m_\gamma$ interacts with  nuclei embedded in the host crystal. For a macroscopic sample, the TAM projection $m_\gamma$ and the spatially inhomogeneous wave front are not controlling the driven transitions; only the pitch angle plays a role \cite{tw_hydrogenlike}. We consider at first a fixed helicity value $\Lambda$. Towards the end of the Section, we present also results for a superposition or two circularly polarized beams with different $\Lambda$ values. All $12$ $\mathcal{M}1$ transitions, $4$ of each allowed $\Delta m$, are driven at once. Again, the $\mathcal{E}2$ channel of the nuclear transition is neglected due to its much smaller radiative coupling. In addition, in the crystal all dopant configurations are equally probable, leading to an equal distribution over the three possible EFG orientations along the $\SI{180}{\degree}$ bonds. This is related to the underlying cubic lattice structure as illustrated in Fig.~\ref{fig:orientations}. As such, the system has in total three mutually perpendicular quantization axes which are shared each by $1/3$ of the total nuclear population. We therefore investigate the excited state population $\rho_{\mathrm{ex},j}$ in a system with multiple quantization axes.

The excited substate populations $\rho_{\mathrm{ex},j}$ are determined via the multilevel Bloch equations in the low saturation limit. In the first place, we calculate a general expression for the matrix element involving an arbitrary orientation of the quantization axis. Then, we solve the multilevel Bloch equations. 
Taking into account all three quantization axes, the occupation of a certain magnetic substate can be written as
\begin{equation}
 \label{eq:totalpop}
  \rho_{\mathrm{ex},j} = \frac{\rho^{q_1}_{\mathrm{ex},j}+\rho^{q_2}_{\mathrm{ex},j}+\rho^{q_3}_{\mathrm{ex},j}}{3}
\end{equation}
where the superscript corresponds to the different orientations of the quantization axis (see Fig.~\ref{fig:orientations}).

In order to account simultaneously for all three orientations of the quantization axis, we need to set first our laboratory frame. 
We choose the latter as depicted in Fig.~\ref{fig:orientations}, with the $z$ axis parallel to $k_z$, i.e., to the propagation direction of the twisted field. 
Additionally, we require a rotation of the nuclear states according to \cite{varshalovich1988quantum}
\begin{equation}
   \begin{aligned}
        \ket{I_g m_g}_n &= \sum_{m_g'} D^{I_g}_{m_g,m_g'}(\alpha, \beta, \gamma) \ket{I_g m_g'}_\ell \\
        &= e^{-im_g \alpha} \sum_{m_g'} d^{I_g}_{m_g,m_g'}(\beta) e^{-im_g' \gamma} \ket{I_g m_g'}_\ell
   \end{aligned}
\end{equation}
where $\alpha =\beta =\gamma =0$ corresponds to the case where $k_z$ and the orientation of the EFG are parallel, see Eq.~\eqref{eq:matrixFinal}.
Here, the subscript $n$ corresponds to the angular momentum eigenstates in the nuclear reference frame and $\ell$ to the light field reference frame, respectively.
The transformation of the sets $\ket{I_e m_e}$ works in a similar fashion.
By following the steps in Sec.~\ref{sec:theory} for $L=1$ and $\mu=\mathcal{M}$ with the given state convention and applying some angular momentum algebra, we arrive at
\begin{equation}
    \begin{aligned}
        M_{fi}^{(\mathrm{tw})} &= -\sqrt{\frac{8\pi \mathcal{I}}{9c\varepsilon_0}}\sqrt{2I_g+1}\sqrt{B(\mathcal{M}1, I_g\rightarrow I_e)} \\
        &\times (-1)^{I_g-m_g}\braket{I_e m_e I_g -m_g|1 \Delta m}  \sum_{M} (-i)^{2 m_\gamma -M}\\ 
& \times e^{i(m_\gamma-M)\phi_b} J_{m_\gamma -M}(\zeta b) d^1_{M\Lambda}(\theta_k)D^1_{M \Delta m}(\alpha,\beta, \gamma).
    \end{aligned}
\end{equation}
The main difference to Eq.~\eqref{eq:matrixFinal} is the additional sum over $M$ and as well as the additional Wigner function $D^1_{M \Delta m}(\alpha,\beta, \gamma)$.
This is in correspondence with the results in \cite{schulz2020generalized} where a similar geometry was investigated for atomic photo-absorption.

In case of a macroscopic sample in which the nuclei are distributed homogeneously over the entire beam, the matrix element modulus squared (related to the observable) must be averaged over a disk of radius $R$ in order to determine the average rate of excitation. This is done via
\begin{equation}
    \ab{\Tilde{M}_{fi}^{\mathrm{(tw)}}}^2 = \frac{1}{R^2 \pi} \int \ab{M_{fi}^{(\mathrm{tw})}}^2 d^2 \bm{b}.
\end{equation}

By using the integral relations \cite{afanasev2013off}
\begin{equation}
\lim \limits_{R \to \infty} \int_0^R J_{m_\gamma-\Delta m}^2(\zeta b) b \, db = \frac{R}{\pi \zeta}
\end{equation}
and
\begin{equation}
    \int_0^{2\pi} e^{i(M'-M)\phi_b} \, d\phi_b = 2 \pi \delta_{M,M'}
\end{equation}
we obtain for the averaged matrix element modulus squared
\begin{equation}
\begin{aligned}
\ab{\Tilde{M}_{fi}^{\mathrm{(tw)}}}^2 &= \ab{\braket{I_e m_e I_g -m_g|1 \Delta m}}^2 B(\mathcal{M}1,I_g \rightarrow I_e) \\
&\times \frac{16\mathcal{I}(2I_g+1)}{9c\varepsilon_0 R\zeta}  \sum_M d^1_{M\Lambda}(\theta_k)^2 d^1_{M \Delta m}(\beta)^2
\end{aligned}
\end{equation}
which is independent of the impact parameter $b$ as well as the Euler angles $\alpha$, $\gamma$ and hence only sensitive to $\beta$ and $\theta_k$.

The optical Bloch equations in the low saturation limit are given by \cite{von2020theory}
\begin{equation}
\label{eq:rhoee}
    \dot{\rho}_{ee} =-i \sum_{g}  \Omega_{eg}\lrb{\rho_{ge}-\rho_{eg}} - \rho_{ee} \Gamma
\end{equation}
for the excited state population and
\begin{equation}
\label{eq:Bloch-coh}
    \dot{\rho}_{ge} = -i \Omega_{eg} \lrb{\rho_{ee}-\rho_{gg}} - \rho_{ge} \Tilde{\Gamma}
\end{equation}
for the coherences.
Here, $\Gamma = \Gamma_\gamma$, since spontaneous EB decay is on the order of $\Gamma_{\text{eb}}\approx \mathcal{O}(\SI{e-8}{\per \second})$ in $^{229}$Th:CaF$_{2}$ \cite{nickerson2020nuclear,bsn_pra} and can be neglected. Once more we neglect relaxation between sublevels of the same manifold. In the Bloch equation \eqref{eq:Bloch-coh}, $\Tilde{\Gamma} \approx \Gamma_\ell/2$ corresponds to the decay rate of the coherences which is dominated by the large linewidth of the laser. A suitable laser source for this procedure can be for instance the system described in Ref.~\cite{thielking2023vacuum}.
The large linewidth of the laser leads to a fast dissipation of the coherences which in turn leads to a rapidly evolving steady state.
In this case, one can set $\dot{\rho}_{ge}=0$. 
In combination with the weak driving field, the ground state population remains basically unaffected such that $\rho_{gg} - \rho_{ee} \approx 1/(2I_g +1)$ holds for an initial ground state quadrupole structure.
The sum in Eq.~\eqref{eq:rhoee} thereby runs over all initial magnetic substates from which the final state can be reached via a $\mathcal{M}1$ transition, i.e. $m_g=m_e - \Delta m$ due to the fact that a vortex beam has $2L+1$ amplitudes. Thus, we can replace $\Omega_{eg}$ by $\Omega_{eg}^{(\text{tw})}$.

With these assumptions, solving Eq.~\eqref{eq:rhoee} with $\rho_{ee}(0)=0$ provides for the nuclear ensemble excitation in the wavefront
\begin{equation}
       \begin{aligned}
        \rho_{ee}^{(\mathrm{tw})} &=  \overbrace{\frac{2}{\Gamma \Tilde{\Gamma}} \frac{1-e^{-\Gamma t}}{2I_g+1}  }^{\alpha(t)}  \sum_g \lrb{\Omega_{eg}^{(\mathrm{tw})}}^2 \\   
        &= \frac{\alpha(t)}{R \zeta} \frac{16 \mathcal{I}(2I_g+1)}{9 c \varepsilon_0 \hbar^2} B(\mathcal{M}1, I_g \rightarrow I_e) \\
       & \times \sum_{m_g,M} \ab{\braket{I_e m_e I_g -m_g|1 \Delta m}}^2  d^1_{M\Lambda}(\theta_k)^2 d^1_{M \Delta m}(\beta)^2.
   \end{aligned}
\end{equation}
In the following, we will refer to the excited magnetic substates as $\rho_{\mathrm{ex},j}^{(\mathrm{tw})}$ instead of $\rho_{ee}^{(\mathrm{tw})}$ as introduced in the labels in Fig.~\ref{fig:manyScenario}.

Summing over all four excited substates we obtain
\begin{equation}
  \begin{aligned}
         {\rho}_{\text{ex}}^{(\mathrm{tw})} &=\sum_{j}    {\rho}_{\mathrm{ex},j}^{(\mathrm{tw})} \\
         &= \frac{\alpha(t)}{R \zeta} \frac{16 \mathcal{I}(2I_g+1)}{9 c \varepsilon_0 \hbar^2} B(\mathcal{M}1, I_g \rightarrow I_e) \, ,
  \end{aligned}
\end{equation}
where the Clebsch-Gordan coefficients and the Wigner small-$d$  functions vanish due to their unitarity.
Then, the fraction of an excited magnetic substate driven by a circularly polarized vortex field is given in the low saturation limit by
\begin{equation}
\begin{aligned}
\label{eq:twcrosscirc} 
&    \frac{\rho_{\mathrm{ex},j}^{(\mathrm{tw})} }{\rho_{\text{ex}}^{(\mathrm{tw})}} =\\
        & \sum_{m_g,M} \ab{\braket{I_e m_e I_g -m_g|1 \Delta m}}^2  d^1_{M\Lambda}(\theta_k)^2 d^1_{M \Delta m}(\beta)^2.
\end{aligned}
\end{equation}

Apart from circularly polarized Bessel beams, it has also been discussed in literature that different beam polarizations can be generated via superpositions of twisted light beams \cite{schulz2020generalized, ramakrishna2022photoexcitation}. In the following  we also consider the superposition of two Bessel beams with the same TAM projection $m_\gamma$ and opposite helicities $\Lambda$ of the form
\begin{equation}
\label{eq:superposA}
    \bm{A}^{(\text{tw})}_{\text{sup}} =\frac{1}{\sqrt{2}} \lrb{\bm{A}_{m_\gamma,\Lambda = +1}^{(\text{tw})} + e^{i \varphi}\bm{A}_{m_\gamma,\Lambda = -1}^{(\text{tw})}  }
\end{equation}
where $\varphi$ is an arbitrary phase. In the paraxial regime and the special case $m_\gamma =0$, this superposition would correspond to a radially polarized vortex field for $\varphi =0$ and an azimuthally polarized vortex field for $\varphi = \pi$ \cite{schulz2020generalized}.

Following the aforementioned steps in the derivation of the matrix element and sublevel population, the partial excited states for such a superposition can be written as
\begin{equation}
\label{eq:lincross}
\begin{aligned}
        \frac{\rho_{\mathrm{ex},j}^{(\mathrm{tw})} }{\rho_{\text{ex}}^{(\mathrm{tw})}}  &=\frac{1}{2}\sum_{m_g,M} \ab{\braket{I_e m_e I_g -m_g|1 \Delta m}}^2 \\
     &\times d^1_{M \Delta m}(\beta)^2 \Big [d^1_{ M \Lambda}(\theta_k)^2 +d^1_{M-\Lambda}(\theta_k)^2\\
     &+ 2 \cos{(\varphi)}d^1_{ M -\Lambda}(\theta_k)d^1_{ M \Lambda}(\theta_k) \Big].
\end{aligned}
\end{equation}

Since both expressions  \eqref{eq:twcrosscirc} and \eqref{eq:lincross} are independent  of $\alpha$ and $\gamma$, the contribution of the misaligned axes are equal such that Eq.~\eqref{eq:totalpop} simplifies to
\begin{equation}
\label{eq:totpop2}
    \rho_{\mathrm{ex},j} = \frac{\rho_{\mathrm{ex},j}(\beta = 0^\circ)+2\rho_{\mathrm{ex},j}(\beta = 90^\circ)}{3}.
\end{equation}

\begin{figure*}
    \centering
    \includegraphics[width = 16cm]{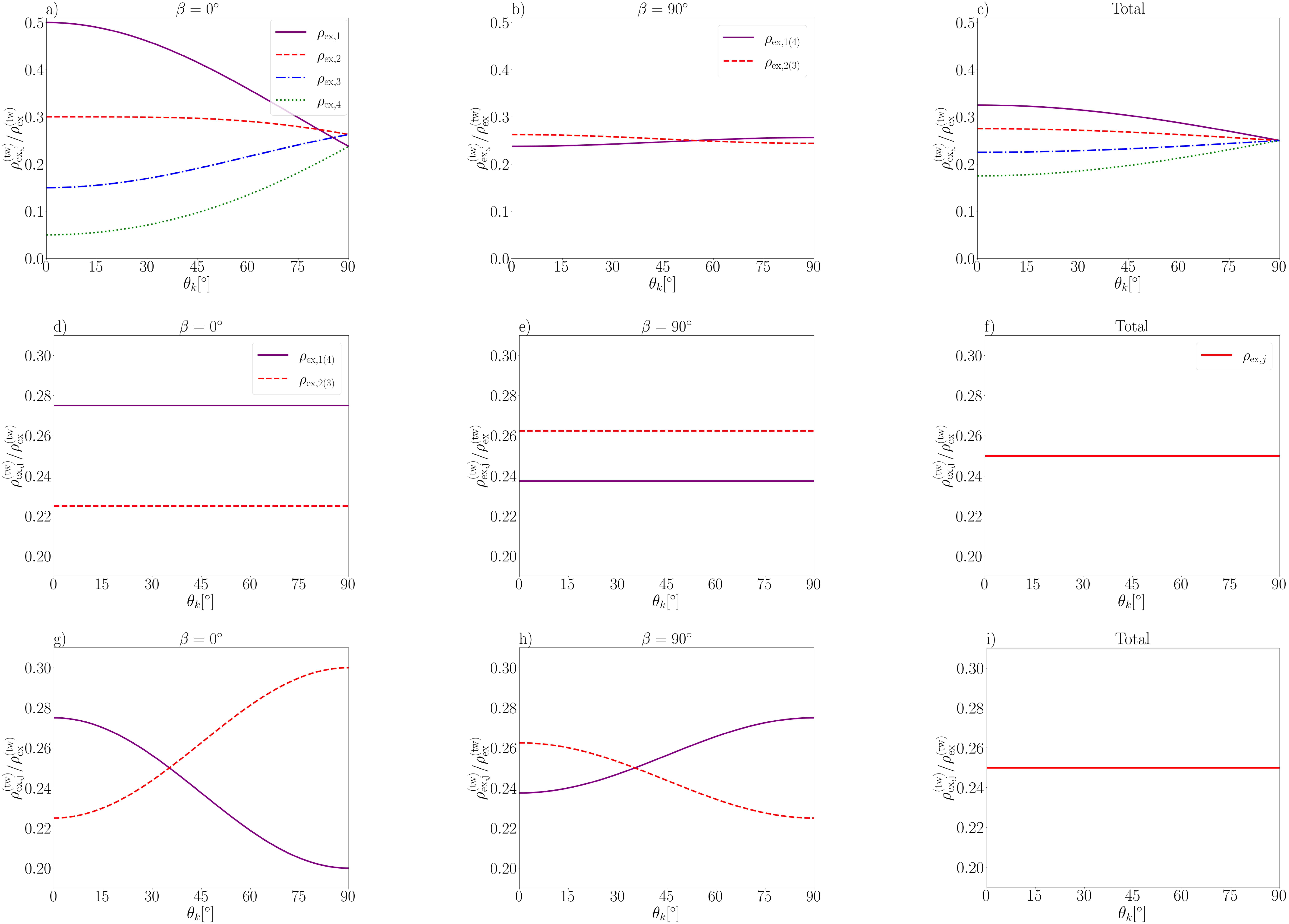}
    \caption{Distribution of magnetic sublevel population as a function of pitch angle for different vortex beam polarizations and crystal quantization axes. a)-c) correspond to a circularly polarized Bessel beam.  a) $\beta = 0$, b) $\beta=\SI{90}{\degree}$ and c) the sum  \eqref{eq:totalpop}. d)-f) correspond to a beam superposition \eqref{eq:superposA} with $\varphi=0$ and g)-i) correspond to a beam superposition \eqref{eq:superposA} with $\varphi=\pi$. The  quantization axes are column-wise the same as in  a)-c). }
    \label{fig:polarization}
\end{figure*}

In Fig.~\ref{fig:polarization}, the partial populations $ \rho_{\mathrm{ex},j}^{(\mathrm{tw})}/\rho_{\text{ex}}^{(\mathrm{tw})}$ as function of the pitch angle are presented for a circularly polarized Bessel beam and for the two-beam superpositions following Eq.~\eqref{eq:lincross} with $\varphi=0$ and $\varphi=\pi$. The scenarios $\beta=0^\circ$, $\beta=90^\circ$ and their combination in Eq.~\eqref{eq:totpop2} are discussed. We choose the pitch angle range $[0^\circ, 90^\circ]$. In practice, the value $\theta_k = \SI{90}{\degree}$ is not feasible experimentally, since this corresponds to a propagating wave without a longitudinal component $k_z$ which is unphysical.
Furthermore, most current experiments with twisted light beams are in the paraxial regime such that an excitation pattern significantly different from plane wave excitation is challenging at the moment. Nevertheless, according to Ref.~\cite{tw_hydrogenlike}, pitch angles in the range $\SI{20}{\degree}$ to $\SI{60}{\degree}$ seem to be experimentally feasible.

%%%%%%%%%%%%%%%%%%%%%%%%%%%%%%%%%%%%%%%%%
%%%%%%%%%%%%%%%%%%%%%%%%%%%%%%%%%%%%%%%%

Figures \ref{fig:polarization}a)-c) represent the results for a circularly polarized Bessel beam with $\Lambda= +1$, starting with a), which corresponds to $\beta=0^\circ$. At small opening angles the distribution hardly deviates from the plane wave result for a circularly polarized beam. Here, the occupation probability is dominated by the underlying Clebsch-Gordan coefficient $|\braket{I_e m_e I_g -m_g|11}|^2$ of the transition. With increasing  $\theta_k$, a change in the sublevel distribution becomes visible. The two magnetic substates $\rho_{\mathrm{ex},1}$ and $\rho_{\mathrm{ex},2}$ with the largest occupation probability become less populated,  while the populations of $\rho_{\mathrm{ex},3}$ and $\rho_{\mathrm{ex},4}$ start to increase. This is related to the fact that the Wigner functions in Eq.~\eqref{eq:twcrosscirc} for $\Delta m = 0$ and $\Delta m=-1$ are  increasing with $\theta_k$.
This progression takes place until the occupation of magnetic substates with the same projection quantum number $\ab{m_e}$ becomes equal and approaches the value
\begin{equation}
\label{eq:supsup}
      \frac{\rho_{\mathrm{ex},j}^{(\mathrm{tw})} }{\rho_{\text{ex}}^{(\mathrm{tw})}} = \frac{3}{16} + \frac{1}{4}\ab{\braket{I_e m_e I_g -m_g|1 0}}^2
\end{equation}
at $\theta_k = \SI{90}{\degree}$.

In Fig.~\ref{fig:polarization}b), we present our results once the quantization axis is rotated by \SI{90}{\degree}. For small values of $\theta_k$, the occupation follows Eq.~\eqref{eq:supsup}. With increasing pitch angle, the distribution changes only slightly. The corresponding results accounting for all three quantization axis via Eq.~\eqref{eq:totalpop} are displayed in Fig.~\ref{fig:polarization}c).
 For small opening angles, the occupation probability follows the pattern visible in part a), however with a different magnitude.
One can observe that states with the same absolute value $\ab{m_e}$ are symmetrically distributed around $1/4$ corresponding to a homogeneous sublevel population distribution throughout the entire range of $\theta_k$. Thus, in the context of NQRS, the case of most interest would be the small pitch angle range for $\beta=0^\circ$ quantization, where the relevant populations  $ \rho_{\mathrm{ex},2}+ \rho_{\mathrm{ex},3}$ and $ \rho_{\mathrm{ex},1}+ \rho_{\mathrm{ex},4}$ differ  by $\approx 0.1$. However, it is not clear whether any means to favor this particular quantization axis in the sample preparation process could be developed.

Figures \ref{fig:polarization}d)-f)  show the results for a vortex beam superposition with relative phase $\varphi=0$. For all quantization axes, the excitation probabilities are independent of the pitch angle $\theta_k$ and are the same as plane wave excitation. 
This is due to the unitarity relation of the Wigner small-$d$  functions. 
For $\beta=0^\circ$ the excitation probability illustrated in Fig.~\ref{fig:polarization}d) is given by 
\begin{equation}
\label{eq:linear}
\begin{aligned}
         \frac{\rho_{\mathrm{ex},j}^{(\mathrm{tw})} }{\rho_{\text{ex}}^{(\mathrm{tw})}} &=\frac{1}{2}\Big (\ab{\braket{I_e m_{e_j} I_g -m_{g_j}|11}}^2 \\
         &+\ab{\braket{I_e m_{e_k} I_g -m_{g_k}|1-1}}^2 \Big)
\end{aligned}
\end{equation}
throughout the entire range of $\theta_k$.  For the $\beta = \SI{90}{\degree}$ case illustrated in Fig.~\ref{fig:polarization}e), the situation is similar, however with the difference that the sublevel distribution follows Eq.~\eqref{eq:supsup} throughout the entire range of $\theta_k$. This is again related to the unitarity relation of the  Wigner function for $L=1$ and $M \neq 0$ and additionally through the misaligned quantization axis leading to an admixture of $\Delta m = 0$ transitions. 
When including all three axes in Fig.~\ref{fig:polarization}f),  all magnetic substates are equally populated over the entire range of $\theta_k$. This could also be achieved by a plane wave field with a linearly polarized beam \cite{von2020theory}.

Finally,  Figs.~\ref{fig:polarization}g)-i) present the calculated excitation probabilities for a superposition of vortex beams with relative phase $\varphi=\pi$. For the case with $\beta=0$ in  Fig.~\ref{fig:polarization}g) the excited state population is determined for small opening angles by   Eq.~\eqref{eq:linear} while later on for  $\theta_k=\SI{90}{\degree}$ it is proportional to the remaining Clebsch-Gordan coefficient  $|\braket{I_e m_e I_g -m_g|1 0}|^2$. 
Similarly, in Fig.~\ref{fig:polarization}h) the population  is proportional to the expression in Eq.~\eqref{eq:supsup} for small angles, and follows   Eq.~\eqref{eq:linear} for large pitch angles. Taking into account all three quantization axes the magnetic substates are equally occupied  throughout the entire range of $\theta_k$.

Summarizing, depending on the twisted light field mode under consideration the polarization of a $^{229}$Th:CaF$_2$ crystal  can be slightly modified compared to the plane wave case by varying the transverse momentum $\zeta$ of a broadband vortex beam. This effect is most pronounced along a single quantization axis in the sample, while taking into account all three intrinsic crystal axes leads to a polarization pattern rather similar to the plane wave result.  The most interesting result is obtained for a circularly polarized vortex beam. The largest excited state polarization is achieved for small pitch angles, thus closer to the plane wave result, however only for the single-quantization axis aligned with the direction of pulse propagation with $\beta=0^\circ$; averaged over all three quantization axis the polarization vanishes.  We note that it is unclear how the case of a single quantization axis could be implemented in practice.

Our results for $^{229}$Th:CaF$_2$ are limited to the $180^\circ$ interstitial fluorine ions configuration. The $90^\circ$ one is more complicated and could offer other possibilities for excited state polarization. Also, other VUV-transparent crystals with a more complicated structure than CaF$_2$ might also offer the possibility to work with a single quantization axis and different quadrupole splitting level schemes.

%----------------- continue here! --------------------------------------------

\section{Conclusion \& Outlook}
\label{sec:fin}
%%%%%%%%%%%%%%%%%%%%%%%%%%%%%%%%%%%%%%%%%%%%%%%%%%%%%%%%%%%%%%%%%%

Nuclear photoabsorption in $^{229}$Th using twisted light was investigated theoretically within a semiclassical approach, employing a classical description of the electromagnetic field. Two scenarios were investigated. 
 First, we focused on the temporal and spatial excitation dynamics of a single $^{229}$Th ion interacting with a resonant vortex beam. Thereby, it was shown that driving of the $\mathcal{E}2$ transition  can be optimized in the center of the vortex beam, yet under at present rather challenging experimental parameters. Adjacent to the beam center, the $\mathcal{E}2$ channel becomes negligible and $\mathcal{M}1$ transitions dominate. We have presented numerical results for 
 nuclear clock transitions between the hyperfine-split levels of $^{229}$Th$^{3+}$ driven by narrow-band vortex beams and discussed their spatial and temporal excitation patterns. 

Second, we have studied the interaction of broad-band twisted light with a macroscopic $^{229}$Th:CaF$_2$ crystal. Here
 all $\mathcal{M}1$ transitions were addressed by the field at once, and the excited state polarization of the crystal was deduced taking into account all three orientations of the quadrupole splitting quantization axis. We have investigated the sublevel population as a function of the beam pitch angle. Our results show that for practical NQRS applications one would require a single aligned EFG quantization axis, in the  paraxial limit of  small $\theta_k$. As the quantization axis plays an important role for the excited state population, our study should be extended to the second relevant EFG configuration in $^{229}$Th:CaF$_2$ and perhaps also to different VUV-transparent crystals which could offer completely different hyperfine splitting features, such as LiCAF and LiSAF. On a different front, our results could be extended to  Laguerre-Gaussian modes to better predict experimentally available vortex beams, and to take into account the coherent pulse propagation through macroscopic samples. This will be the subject of future work.

\begin{acknowledgments}
%%%%%%%%%%%%%%%%%%%%%%%%%%%%%%%%%%%%%%%%%%%%%%%%%%%%%%%%%%%%%%%%%%%%%%%%%%
AP gratefully acknowledges support from the Deutsche Forschungsgemeinschaft (DFG, German Science Foundation) in the framework of the Heisenberg Program (PA 2508/3-1). The research was supported by the Austrian Science Fund (FWF) [grant DOI:10.55776/F1004] (COMB.AT) together with the DFG (PA 2508/5-1). This work is also part of the ThoriumNuclearClock project that has received funding from the European
Research Council (ERC) under the European Union’s Horizon
2020 research and innovation programme (Grant Agreement
No. 856415).

\end{acknowledgments}

\bibliography{refs}

\end{document}